\pdfoutput=1
\documentclass[11pt]{article}
\usepackage{jheppub}

\usepackage{amsmath} 
\usepackage{multirow} 
\usepackage{amsfonts}
\usepackage{amssymb,graphics,psfrag}
\usepackage{array,epsfig,multirow,graphicx,mathtools}
\usepackage{hyperref}
\usepackage{xspace}
\newcommand{\beq}{\begin{equation}}
\newcommand{\eeq}{\end{equation}}
\newcommand{\be}{\begin{equation}}
\newcommand{\ee}{\end{equation}}
\newcommand{\bea}{\begin{eqnarray}}
\newcommand{\eea}{\end{eqnarray}}   
\newcommand{\ben}{\begin{eqnarray*}}
\newcommand{\een}{\end{eqnarray*}}                  
\newcommand{\ba}{\begin{aligned}}
\newcommand{\ea}{\end{aligned}}
\newcommand{\bt}{\begin{tabular}}
\newcommand{\et}{\end{tabular}}
\newcommand{\bc}{\begin{center}}
\newcommand{\ec}{\end{center}}

\newcommand{\cO}{\mathcal{O}}

\newcommand{\cE}{\mathcal{E}}

\newcommand{\cK}{\mathcal{K}}
\newcommand{\cN}{\mathcal{N}}

\newcommand{\cA}{\mathcal{A}}

\newcommand{\cF}{\mathcal{F}}

\newcommand{\cV}{\mathcal{V}}

\newcommand{\cM}{\mathcal M}

\newcommand{\I}{\text{Im}}

\newcommand{\bbZ}{\mathbb{Z}}

\newcommand{\nn}{\nonumber}

\newcommand{\cref}{{\bf [check ref]}}

\def\a{$\alpha'$\xspace}
\def\g{$g_s$\xspace}
\def\s{$SL(2,\mathbb{Z})$\xspace}

\newcommand{\Li}{{\rm Li}}
\psfrag{n1}{$\nu_1$}
\psfrag{n2}{$\nu_2$}
\psfrag{n1'}{$\nu_1'$}
\psfrag{n2'}{$\nu_2'$}
\psfrag{n9}{$\nu_9$}
\psfrag{n10'}{$\nu_{10}'$}
\psfrag{t1}{$\tilde{\nu}_1$}
\psfrag{t2}{$\tilde{\nu}_2$}
\psfrag{t9}{$\tilde{\nu}_9$}
\psfrag{t1'}{$\tilde{\nu}_1'$}
\psfrag{t2'}{$\tilde{\nu}_2'$}
\psfrag{t10'}{$\tilde{\nu}_{10}'$}

\newcommand{\Ip}{I$'$\xspace}

\title{On quantum corrected K\"ahler potentials in F-theory}

\author[a]{I\~naki Garc\'ia-Etxebarria,}
\author[b]{Hirotaka Hayashi,}
\author[c]{Raffaele Savelli,}
\author[d,e]{and Gary Shiu}

\affiliation[a]{Theory Group, Physics Department, CERN, CH-1211,
  Geneva 23, Switzerland}

\affiliation[b]{School of Physics, Korea Institute for Advanced Study,
  Seoul 130-722, Korea}

\affiliation[c]{Max-Planck-Institut f\"ur Physik, F\"ohringer Ring 6,
  80805 Munich, Germany}

\affiliation[d]{Department of Physics, University of
  Wisconsin-Madison, Madison, WI 53706, USA}

\affiliation[e]{Department of Physics and Institute for Advanced
  Study, Hong Kong University of Science and Technology, Hong Kong}

\emailAdd{inaki@cern.ch}
\emailAdd{hayashi@kias.re.kr}
\emailAdd{savelli@mpp.mpg.de}
\emailAdd{shiu@physics.wisc.edu}

\abstract{We work out the exact in \g and perturbatively exact in \a
  result for the vector multiplet moduli K\"ahler potential in a
  specific $\cN=2$ compactification of F-theory. The well-known
  $\alpha'^3$ correction is absent, but there is a rich structure of
  corrections at all even orders in \a. Moreover, each of these orders
  independently displays an \s invariant set of corrections in the
  string coupling constant. This generalizes earlier findings to the
  case of a non-trivial elliptic fibration. Our results pave the way
  for the analysis of quantum corrections in the more complicated
  $\cN=1$ context, and may have interesting implications for the study
  of moduli stabilization in string theory.}

\begin{document}

\makeatletter
\let\old@fpheader\@fpheader
\renewcommand{\@fpheader}{\begin{flushright}\old@fpheader\hfill 
CERN-PH-TH/2012-359, MAD-TH-12-09,\\MPP-2012-185, KIAS P-12085\end{flushright}}
\makeatother

\maketitle

\newpage

%
%

\section{Introduction}

In recent years, there has been a resurgence of interest in F-theory
\cite{Vafa:1996xn}.  This renewed interest is due largely to the
observation that certain realistic particle physics features, such as
the gauge group, matter content, and couplings of Grand Unified
Theories (GUTs), can be elegantly obtained in this framework
\cite{Donagi:2008ca,Beasley:2008dc, Hayashi:2008ba, Beasley:2008kw} (see \cite{Weigand:2010wm,Maharana:2012tu} for recent reviews).
Besides extending the D-brane phenomenology program to describe
realistic GUTs, another (perhaps the original) appeal of F-theory is
that it provides a geometrical way to formulate and analyze type IIB
string vacua non-perturbatively.

The power of F-theory lies in its potential to geometrically describe
the non-perturbative physics of string theory, but aspects of its
effective action obtained so far have not yet fully exploited this
property.  In lack of an action principle or a microscopic formulation
of F-theory, one often has to rely on F-theory as a limit of M-theory
to obtain its low energy effective action
\cite{Dasgupta:1999ss,Denef:2008wq,Grimm:2010ks,Grimm:2011sk,Grimm:2011fx,Bonetti:2011mw,Grimm:2012rg} (see \cite{Becker:1996gj} for the starting M-theory solutions).
While geometry and low energy consistency conditions impose
constraints on the low energy effective action\footnote{In \cite{Marchesano:2008rg, Marchesano:2010bs}, additional constraints from the proper coupling between open and closed strings were used to determine the K\"ahler potential for type IIB theory in the presence of fluxes as well its generalization to F-theory.}, the underlying
symmetries of type IIB string theory are not always apparent.  In this
paper, we shall make extensive use of string dualities in order to
derive aspects of the quantum corrected effective action of F-theory.
In particular, we explore simple F-theory models which admit several
dual descriptions (see figure \ref{fig:duality} for the web of string
dualities involved).  The dual descriptions enable one to compute in
some cases not only the perturbative $\alpha'$ corrections to
F-theory, but also results that are fully non-perturbative in the
string coupling. In our approach, the non-perturbative $SL(2,{\mathbb
  Z})$ symmetry is manifest in the effective action. The K\"ahler
potential so obtained should generalize fully non-perturbatively the
type IIB result.

One of the interesting features of F-theory compactifications is that
one may be able to naturally combine phenomenological model
construction with moduli stabilization analysis \cite{Blumenhagen:2008zz, Collinucci:2008sq, Cicoli:2011qg}. Given that the
leading $\alpha'$ corrections to the K\"ahler potential has played a
key role in the so-called LARGE Volume Scenario (LVS) of
\cite{Balasubramanian:2005zx, Conlon:2005ki}, we expect that our generalization of
these type IIB results to F-theory should have some interesting
implications to moduli stabilization. In LVS, string corrections to
the tree-level supergravity effective action computed in
\cite{Becker:2002nn} play an essential role, and a volume modulus is
stabilized so that the compactification volume is as large as
$10^{15}$ in string units. Since the scenario relies on the specific
string correction of $\mathcal{O}(\alpha'^3)$ in the string frame to
the K\"ahler potential, other corrections might have some effects on
the moduli stabilization. Indeed, some perturbative one-loop $g_s$ corrections to some $\mathcal{N}=1$ and
$\mathcal{N}=2$ toroidal orientifold models were computed in \cite{Berg:2005ja} (see also \cite{Antoniadis:1996vw}). Ref.~\cite{Berg:2005ja} found corrections of
order $\mathcal{O}(g_s^2\alpha'^2)$ in the string frame to the
K\"ahler potential.\footnote{The parametric form of these loop corrections was earlier found in \cite{vonGersdorff:2005bf}. A similar set of corrections were found
from the heterotic perspective in \cite{Anguelova:2010ed}.} Hence,
one has to check which corrections are leading, in order to find a
true minimum. Interestingly, there can be certain cancellations for
the latter correction in the scalar potential, and some models are
robust against the inclusion of the latter correction in a certain region of the moduli space \cite{vonGersdorff:2005bf, Berg:2005yu, Berg:2007wt, Cicoli:2007xp}. In this respect, one of our aims here is to generalize the
result in \cite{Berg:2005ja} to full $g_s$ corrections including
non-perturbative terms in $g_s$. This is particularly important for
F-theory compactifications since typical GUTs require a strong
coupling effect in $g_s$ for generating some favorable
phenomenological features. Although $\mathcal{N}=1$ models are of
interest for this purpose, $\mathcal{N}=2$ models will still exhibit
interesting structures in the corrections. In fact, the qualitative
features of $\mathcal{N}=2$ corrections is similar to the
$\mathcal{N}=1$ corrections in toroidal orientifold models considered
in \cite{Berg:2005ja}. As mentioned before, on the other hand,
$\mathcal{N}=2$ supersymmetry is powerful enough to obtain fully
non-perturbative $g_s$ corrections as well as all the perturbative
$\alpha'$ corrections. Motivated by these observations, we will work
on finding the effective action of a particular $\mathcal{N}=2$
F-theory model in this paper.

More precisely, we concentrate on F-theory compactified on
K3$\times$K3. We shall be able to disentangle \g and \a corrections
and discuss the roles played by the various moduli of the two K3
manifolds. In particular, K\"ahler modulus and complex structure
moduli of the elliptic K3, while decoupled at tree-level in \a, are
non-trivially mixed at loop-level. The structure of this mixing is
rigidly constrained by the \s invariance of the underlying type IIB
theory and we will propose a purely F-theoretic interpretation of this
fact coming from the M-theory description of F-theory, generalizing the
results for compactifications on trivial elliptic fibrations first
found in \cite{Green:1997tv,Green:1997di,Green:1997as,Green:1998by}.

This paper is organized as follows. In section 2, we discuss some
general issues about the effective action of F-theory, and set up the
computation of the K\"ahler potential which we aim to address in this
work.  In section 3, we review the basics of the string theory model
under consideration, focusing on how various supersymmetry multiplets
transform and on the duality relations connecting the type IIB model
to type I and to heterotic. In section 4, we systematically analyze
the threshold corrections to the K\"ahler metric of the vector
multiplet moduli space of type \Ip string (and hence F-)theory, both
with and without Wilson lines. We will check explicitly that the
K\"ahler potential to each perturbative order in $\alpha'$ is
invariant under an $SL(2,{\mathbb Z})$ symmetry.  In section 5, we
provide a geometric, F-theory interpretation of the type IIB result in
section 4 by making use of the F/M-theory duality. We shall argue that
the threshold corrections to the K\"ahler potential can be interpreted
in F-theory as coming from integrating loops of 11D super gravitons
with various momenta.  We conclude in section 6. Some important but
more technical details are relegated to the appendices.

\section{Setting up the problem}

We would like to take some steps towards understanding quantum
corrections to the K\"ahler potential of F-theory
compactifications. In particular, F-theory already represents a
completion of type IIB string theory as far as string-loop corrections
are concerned, but it is perturbative with respect to $\alpha'$
corrections, exactly on the same footing as type IIB supergravity. This fact is
reflected in the basic objects of the effective field theory arising
from an F-theory compactification
\cite{Grimm:2010ks,Denef:2008wq}. For instance, consider F-theory
compactified on a smooth, elliptically fibered Calabi-Yau fourfold. At
tree level in \a, the 4d, $\mathcal{N}=1$ K\"ahler potential splits in
two decoupled contributions:
\begin{align}\label{KkKc}
\mathcal{K}=\mathcal{K}_K+\mathcal{K}_c\,,
\end{align}
where the first is due to moduli of the K\"ahler structure only and
the second to moduli of the complex structure only of the internal
fourfold. Explicitly they look like:
\begin{align}\label{KaehlerPot}
  \mathcal{K}_K=-3\log\mathcal{V}_{{\rm
      CY}_4}\,,\qquad\qquad\mathcal{K}_c=-\log\int_{{\rm
      CY}_4}\Omega_4\wedge\bar\Omega_4\,,
\end{align}
where $\cV_{{\rm CY}_4}$
is the classical volume of the Calabi-Yau fourfold, while $\Omega_4$
is its unique holomorphic $(4,0)$-form. The complex structure moduli
of the internal fourfold contain three different kinds of moduli of
the underlying type IIB weak coupling orientifold compactification
(Sen limit): The bulk moduli of the internal Calabi-Yau threefold
(closed string moduli), the 7-brane deformation moduli (open string
moduli)\footnote{The separation between bulk and brane-type moduli is
  not canonical, but for our illustrative purposes it is not needed to
  go into the details of this subtlety.} and the axio-dilaton
$S=C_0+ie^{-\phi}$, thought of as an actual 4d modulus. Indeed,
generically the complex structure of the torus fiber is not a modulus
because it varies over the internal space according to the implicit
relation \be\label{KleinFunction} j(S({\bf
  z}))=\frac{4(24f)^3}{27g^2+4f^3}({\bf z})\,, \ee where $j$ is the
modular invariant Klein function while $f$ and $g$ are polynomial
functions of the base coordinates ${\bf z}$, defining the Weierstrass
representation of the elliptic fibration. The solution for $S$ of
eq. \eqref{KleinFunction} encodes the backreaction on the axiodilaton
of a given 7-brane solution of type IIB string theory.  However, the
Sen parameterization of $f$ and $g$ allows to isolate from the
backreacted solution a constant piece $S_0$, which represents the
asymptotic value of the axiodilaton far away from the 7-brane sources
in a given chart\footnote{In Sen's limit all 7-brane sources are mutually local, and one can always choose the frame where they are D7-branes. Consequently, there will be no monodromies affecting the dilaton.}, and thus behaves as a true 4d modulus.

In general the computation of the periods of $\Omega_4$ to evaluate
$\mathcal{K}_c$ is extremely hard, and possible only in case one has
few moduli. However, to make clear our purposes, it is instructive to
consider its weak coupling limit. Taking the Sen limit of an F-theory
compactification just means finding a region in the complex structure
moduli space of the fourfold in which the imaginary part of the
axio-dilaton can be sent to infinity in a globally well-defined
way. In doing so one sees that the discriminant of the elliptic
fibration gets factorized in two pieces, whose vanishing locus can be
interpreted in a suitable $SL(2,\mathbb{Z})$ frame as a D7-brane and
an O7-plane. Since now the string coupling constant $g_s$ can be kept
small everywhere on the base (except on the locus where the O7 sits),
one can make a perturbative expansion\footnote{One can also obtain in
  the Sen limit a complete perturbative expression, which is only up
  to purely non-perturbative terms going like $e^{-1/g_s}$
  \cite{Denef:2008wq}.} of $\mathcal{K}_c$ in $g_s$:
\begin{align}\label{FTheoryTreealpha}
  \mathcal{K}_c=-\log({\rm Im} S_0)-\log i\int_{{\rm
      CY}_3}\Omega_3\wedge\bar\Omega_3+\frac{g_s}{2i\int_{{\rm
        CY}_3}\Omega_3\wedge\bar\Omega_3}\mathcal{K}_{\rm
    D7}+\mathcal{O}(g_s^2)\,,
\end{align}
where the first two terms are respectively the standard K\"ahler
potentials for the dilaton and for the complex structure moduli of the
CY threefold in type IIB string theory. The third term governs the
D7-brane moduli and it depends on both open and closed string
moduli. Notice that it enters at linear order in $g_s$, which means
that the backreaction of the D7-branes on the bulk geometry is
suppressed by a power of $g_s$. Therefore at lowest order in $g_s$ no
open string moduli appear at all.

From the analysis above one therefore expects that the full
$\mathcal{K}_c$ in eq. \eqref{KaehlerPot} contains all the $g_s$
corrections of type IIB string theory, perturbative or not. Moreover,
since $\mathcal{K}_c$ only depends on the fibration structure of the
fourfold, one also expects that the whole set of corrections appears
in it in an $SL(2,\mathbb{Z})$ invariant fashion\footnote{We mean here
  that all the physical quantities, like the K\"ahler metric, should
  be invariant.} for the complex structure of the fiber. Indeed, in
any point of the moduli space, if one applies an overall
$SL(2,\mathbb{Z})$ transformation to the corresponding fourfold one
does not change its intrinsic fibration structure, but rather one is
trivializing each chart of the base in a different way, but all at the
same time, so that the transition functions do not change. In other
words, over each chart of the base, one is taking a different
representative of the complex structure of the torus fiber above that
chart, in such a way that the transitions between two intersecting
charts do not change. Consequently, one changes the names of all the
7-branes which appears (namely the monodromy that defines them), but
their mutual relations are untouched. Of course in the perturbative
expansion just described the $SL(2,\mathbb{Z})$ symmetry is explicitly
broken by a preferred choice of $SL(2,\mathbb{Z})$-frame (in the weak
coupling limit only D7's and O7's appear), which allows us to
consistently retain only a few orders in $g_s$ (neither the monodromy
around a D7 nor the one around an O7 contains the `S' generator of
$SL(2,\mathbb{Z})$). The essence of section \ref{ThresCorr} will be to
use, in a concrete model, powerful results from heterotic string
theory to sum up \emph{all} \g corrections in type IIB for a given \a
order. In doing so, each O7-plane is actually resolved in a couple of
mutually non-perturbative (p,q)7-branes. Nevertheless our focus will
not be on the full backreacted solution $S({\bf z})$, as the latter is
a consequence of the intrinsic structure of the F-theory
fibration. Rather we will concentrate on the 4d modulus $S_0$ and on
its \s-class. To anticipate the result, we will verify that physical
quantities will not depend on the specific representative of that
class at every order in \a. Consequently, the K\"ahler potential will
only be invariant up to K\"ahler transformations and this is due to
the fact that the explicit expression for the K\"ahler potential is
usually written in the covering space of the modulus $S_0$, namely the
upper half complex plane. Hence the K\"ahler transformations are
changes between patches within the K\"ahler moduli space induced by
the \s transition functions acting on the \emph{local} coordinate
$S_0$.\footnote{As we will see in our working model, $S_0$ may not be
  a good K\"ahler coordinate everywhere in the moduli space
  \cite{Berg:2005ja}.}

By viewing F-theory as M-theory on the same fourfold upon sending the volume of the fiber to 0 (F-theory limit), one may suspect that the $SL(2,\mathbb{Z})$ invariance of the K\"ahler potential (up to K\"ahler transformations) only holds when the CY fourfold is trivially fibered (no 7-branes, thus constant axiodilaton, i.e. $S=S_0$). Indeed, in this case ${\rm CY}_4={\rm CY}_3\times T^2$ and $SL(2,\mathbb{Z})$ is now a target space duality of the M-theory background and hence any physical quantity is invariant under this group. This property has been highlighted in the computations of \cite{Collinucci:2009nv}. However, the geometrical, sketchy argument presented above is not restricted to the trivial case and suggests that this invariance property persists in more general cases.

One can also argue the $SL(2, \bbZ)$ invariance at the level of the Weierstrass form at least for a smooth case. At each point in the base of a smooth elliptically fibered Calabi--Yau fourfold, the defining equation with a section may be written by the Weierstrass form
\be
y^2 = x^3 + fx + g,
\ee
where $f$ and $g$ may be expressed as 
\be
f = -15\sum_{\omega \in m\omega_1 + n\omega_2}\frac{1}{\omega^4}, \quad g = -35\sum_{\omega \in m\omega_1 + n\omega_2}\frac{1}{\omega^6}. \label{fg}
\ee
Here $m$ and $n$ in the sum are integers except for $(m,n)=0$, and $\omega_1$ and $\omega_2$ are the two periods of the lattice defining the torus. The complex structure $\tau$ of the torus is related to the periods by $\tau = \frac{\omega_2}{\omega_1}$. Since the sum in \eqref{fg} is taken for all the periods except for $0$, the $SL(2, \bbZ)$ transformation for $\omega_1$ and $\omega_2$
\bea
\omega_{2}^{\prime} &=& a\omega_2 + b\omega_1,\\
\omega_1^{\prime} &=& c\omega_2 + d\omega_1
\eea
with $a,b,c,d \in \bbZ$ and $ad-bc=1$ does not change $f$ and $g$. One can do the same $SL(2, \bbZ)$ transformation at every point in the base of the elliptically fibered Calabi--Yau fourfold. Therefore, the defining equation of the smooth elliptically fibered Calabi--Yau fourfold does not change by the $SL(2, \bbZ)$ transformation. For a singular Calabi--Yau fourfold, we may have 7-branes and also matter fields from the intersection between 7-branes in some singular loci. The gauge fields or matter fields may be realized by string junctions between the 7-branes. The configurations of the string junctions also do not change by the overall $SL(2, \bbZ)$ transformation. 

In order to argue the $SL(2, \bbZ)$ invariance at the level of the K\"ahler potential in a low energy effective field theory, there might be a subtlety if the $SL(2, \bbZ)$ transformation involves a weak--strong coupling transformation. For example, one might not have a local Lagrangian description if the $SL(2,\bbZ)$ transformed theory becomes  strongly coupled. However, the gauge couplings of the gauge fields coming from the Kaluza--Klein reduction or the gauge fields on the D7-branes do not change under the $SL(2, \bbZ)$ transformation. For the Kaluza-Klein gauge fields, the kinetic term arises from the dimensional reduction of the ten-dimensional Einstein-Hilbert action in the Einstein frame. Since the metric in the Einstein frame does not change under the $SL(2, \bbZ)$ transformation, the gauge coupling for the Kaluza-Klein gauge fields does not change. For the gauge fields on D7-branes, the gauge coupling is roughly
\be
\frac{1}{g_{\text{YM}}^2} \sim \frac{\text{Vol(4-cycle)}}{g_s},\label{gauge.D7}
\ee
where the $\text{Vol(4-cycle)}$ stands for the string frame volume of the four-cycle which the D7-branes wrap. The expression \eqref{gauge.D7} becomes in the Einstein frame
\be
\frac{1}{g_{\text{YM}}^2} \sim \frac{\text{Vol(4-cycle)}}{g_s} = \widehat{\text{Vol}}(\text{4-cycle}),
\ee
where $\widehat{\text{Vol}}(\text{4-cycle})$ represents the Einstein frame volume of the four-cycle. Therefore, the gauge coupling for the gauge fields on the D7-branes does not change under the $SL(2, \bbZ)$ transformation. To summarize, the gauge couplings for the two types of the gauge fields remains to be weak after the $SL(2, \bbZ)$ transformation if the original gauge couplings are weak. Hence, one may safely use the K\"ahler potentials on both sides and argue the $SL(2,\bbZ)$ invariance of the corresponding K\"ahler metrics. 

\begin{figure}
  \centering
  \includegraphics[width=0.6\textwidth]{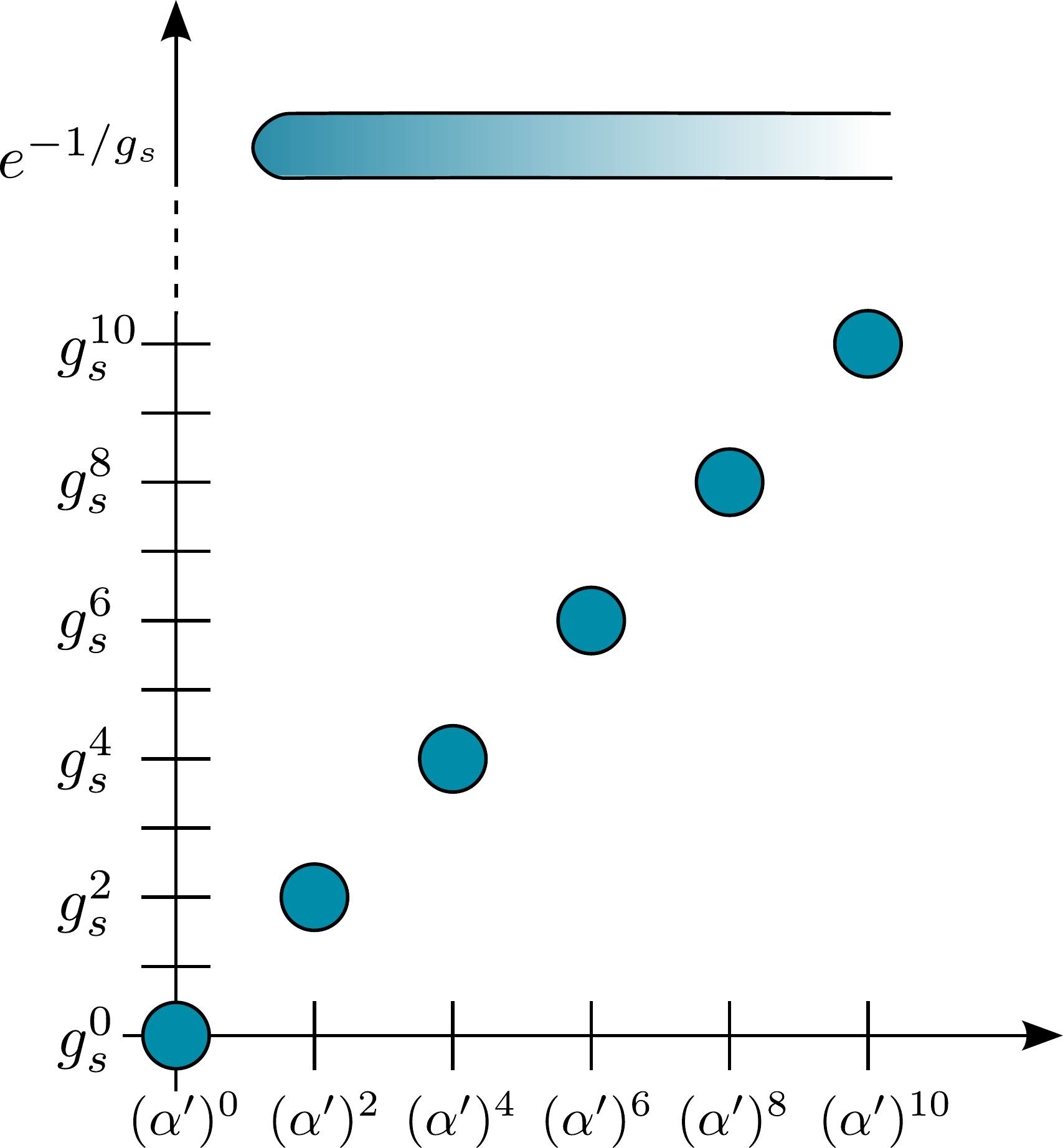}

 \caption{Quantum corrections to the prepotential of F-theory on
    $K3\times K3$. The horizontal axis labels the degree in \a
    of the correction, and the vertical one the degree in \g. The
    circles represent the non-vanishing perturbative terms in the prepotential.
    The solid band on top
    represents the set of non-perturbative corrections in \g;
    Notice that there are no such corrections at tree level in
    \a, as discussed in the text.}
  \label{fig:corrections}
\end{figure}

Moreover we expect that this property still holds at higher order in
\a. Eq. \eqref{KaehlerPot} is only the tree-level expression in \a. At
higher \a-orders, in general, K\"ahler and complex structure moduli
will mix. Nevertheless, since the $SL(2,\mathbb{Z})$ duality of type
IIB string theory is believed to hold at all orders in \a, we expect
that, at each \a-order, there will be a suitable F-theoretic
expression depending on geometrical quantities of the internal CY
fourfold which contains a sum of all $g_s$ corrections in an
$SL(2,\mathbb{Z})$ invariant fashion, as $\mathcal{K}_c$ in
eq. \eqref{KaehlerPot} does for the lowest \a order. In other words,
all kinds of corrections take place in a square (see
fig. \ref{fig:corrections}), in which the horizontal line corresponds
to \a corrections and the vertical to $g_s$ ones. Each \a-tower of
corrections to physical quantities should display the
$SL(2,\mathbb{Z})$ invariance.

In the following we try to verify the above statements by studying a
specific, well known F-theory background, namely F-theory on $K3\times
K3$ \cite{Sethi:1996es,Bershadsky:1997ec,Kakushadze:1998cd,Lerche:1998nx,Greene:2000gh,Gorlich:2004qm,Lust:2005bd,Aspinwall:2005ad,Valandro:2008zg}.  On the one hand this is more general than in
\cite{Collinucci:2009nv} because it involves a non-trivial fibration
(the second $K3$ is elliptically fibered over a 2-sphere). On the
other hand, however, this model is particularly well tractable because
it leads to an $\mathcal{N}=2$ four-dimensional effective theory which
enjoys all the nice non-renormalization theorems for its relevant
quantities. In addition, in this case, we have at our disposal a
well-studied dual heterotic model, in which corrections have been
computed explicitly. This will help us understand systematically
the structure of both \a and \g corrections, which will turn out to be
particularly easy.  We will find all our expectations verified and
provide a clear picture of the whole duality web of corrections,
stating precisely which points of the square in
fig.~\ref{fig:corrections} is occupied by a correction, and what is
the explicit form of the latter. Notice that the simplification
arising in this model is actually a consequence of the $\mathcal{N}=2$
supersymmetry. In $\mathcal{N}=1$ models \a and \g corrections might
be entangled in an intricate way, which may not allow one
to easily isolate the \a-towers of \g corrections and check their \s
invariance.

\section{Review of the model}\label{ReviewModel}

In this section we will review the basics of the string theory model which we want to work with, focusing on the susy representations in which the various low-energy fields transform and on the duality relations connecting the type IIB model to type I and to heterotic. We will not give an extensive treatment, but rather pay attention only to the details we will make use of in the sequel and, in particular, describe the F-theory lift of this model.

\subsection{Generalities}

We consider the so called type I$^\prime$ string theory, namely type
IIB compactified on $K3\times T^2/\mathbb{Z}_2$, where the
$\mathbb{Z}_2$ quotient is an orientifold, whose geometrical action is
just to reflect the coordinates of $T^2$. This action has four fixed
points which are regarded as the positions of four $O7^-$-planes
wrapping $\mathbb{R}^{1,3}\times K3$. This compactification leads to
an $\mathcal{N}=2$, 4d effective theory, which is the orientifold
truncation of an $\mathcal{N}=4$ one. The various moduli fields of the
low-energy theory will arrange into vector multiplets and
hypermultiplets as follows. The complex structure of $T^2$ ($U$), the
overall K\"ahler modulus (volume plus axion) of $K3$ ($T$), the
axio-dilaton ($S$) \footnote{We mean here $S_0$ in the notation
  introduced previously. But to avoid cluttering notations, we will
  drop the subscript throughout the rest of the paper.} and the
transverse positions of the 16 D7-branes ($C^i$) which are needed to
cancel the 7-brane tadpole are all scalar components of 19 vector
multiplets.\footnote{When we do not write a subscript on the moduli
  fields we will always mean quantities in the type \Ip
  theory.} The complex structure moduli of $K3$ plus the K\"ahler
modulus of $T^2$ will instead constitute the scalar fields of a number
of hypermultiplets. Note that this is different from the
compactifications of Type IIB string theory on Calabi-Yau threefolds
where the dilaton is a scalar component of the universal
hypermultiplet. This is because \cite{Tripathy:2002qw} there the
vector fields come from the reduction of RR fields in the usual
Calabi-Yau threefold compactifications. Hence, the gauge kinetic terms
do not have the dilaton dependence. On the other hand, in the present
case, there are gauge fields coming from the reduction of
$B_{2}$. Therefore, there exists a gauge kinetic term of the form
\cite{Andrianopoli:2003jf}
\begin{equation}
e^{-2\phi_{10}^{I^{\prime}}}\sqrt{G_{4}}\sqrt{G_{T^{2}}}\sqrt{G_{K3}}G^{\mu_{1}\nu_{1}}G^{\mu_{2}\nu_{2}}G^{ij}_{T^{2}}H_{\mu_{1}\mu_{2}i}H_{\nu_{1}\nu_{2}j}.
\end{equation}
where $i,j$ are directions of the torus.\footnote{Although $B_{2}$
  itself is odd under the $\mathbb{Z}_{2}$ action, $B_{2}$ with one
  leg on $T^{2}$ is $\mathbb{Z}_{2}$ even.} Then the kinetic term of
the gauge fields contain the dilaton $e^{-2\phi_{10}^{I^{\prime}}}$,
the volume of $K3$, $\sqrt{G_{K3}}$, and the complex structure of
$T^{2}$, $\sqrt{G_{T^{2}}}G^{ij}$. Since the gauge kinetic term in
$\mathcal{N}=2$ supersymmetric field theory is written in terms of
$\mathcal{N}=2$ vector multiplets, we conclude that $S_{I^{\prime}},
T_{I^{\prime}}, U_{I^\prime}$ are scalar components of vector
multiplets.

Our main interest here is to study corrections to the metric of the
vector multiplet moduli space, which is a Special K\"ahler
manifold. Hence all we need is the prepotential as a function of our
19 moduli. Due to its holomorphicity property, quantum corrections to
the prepotential are very well under control and this constitutes an
enormous simplification in carrying out our
analysis.

This type IIB model has also the advantage of admitting a chain of dualities to other type of string theories. Indeed, type I$^\prime$ string theory can be obtained via T-duality from type I compactified on $K3\times T^2$ which in turn is S-dual to heterotic string theory again on $K3\times T^2$. In the next subsection we define the fields we are going to deal with and provide a complete, clear dictionary of this chain of dualities acting on them.

\subsection{Duality dictionary}\label{DualityDictionary}

In this subsection we determine how the \emph{classical} moduli fields of $\mathcal{N}=2$, 4d vector multiplets coming form heterotic string theory compactified on $K3\times T^2$ are related with the ones from type I and type I$^{\prime}$ string theories under the following chain of dualities: 
\begin{table}[h]
\begin{center}
\begin{tabular}{ccccc}
&S-duality&&T-duality&\\
Heterotic &$\longleftrightarrow$& Type I&$\longleftrightarrow$& Type I$^{\prime}$ 
\end{tabular}
\end{center}
\end{table}

\subsubsection{10d duality}

Let us first consider the duality between $SO(32)$ heterotic string theory and type I string theory in ten dimensions. We have the following relations \cite{Witten:1995ex}
\begin{eqnarray}
\phi_{10}^{H} &=& -\phi_{10}^{I},\label{eq:Het-I1}\\
G^{H} &=& e^{-\phi_{10}^{I}}G^{I},\label{eq:Het-I2}
\end{eqnarray}
where $\phi_{10}^{H,(I)}$ is the ten dimensional dilaton and $G_{H,(I)}$ is the metric in heterotic string theory (type I string theory). The relations \eqref{eq:Het-I1}, \eqref{eq:Het-I2} can be derived from the low-energy effective actions of heterotic string theory and type I string theory. The heterotic string effective action in ten dimensions scales with the dilaton $\phi_{10}^{H}$ like
\begin{equation}
\int d^{10}x\sqrt{G_{H}}e^{-2\phi_{10}^{H}}(R_{H}+|\nabla \phi_{10}^{H}|^{2}+F^{2}+|dB|^{2}).\label{eq:hee}
\end{equation}
If we transform \eqref{eq:hee} using \eqref{eq:Het-I1}, \eqref{eq:Het-I2}, the scaling becomes
\begin{equation}
\int d^{10}x \sqrt{G_{I}}(e^{-2\phi_{10}^{I}}(R_{I}+|\nabla \phi_{10}^{I}|^{2}) + e^{-\phi_{10}^{I}}F^{2} + |dC|^{2}).\label{eq:typeI}
\end{equation}
Then, \eqref{eq:typeI} has the correct scaling behavior for the type I string effective action.

\subsubsection{4d duality}

Now we consider the compactification on $T^{2} \times K3$ and see how the S-duality relates the moduli on both sides \cite{Antoniadis:1996vw}. Since the moduli spaces of vector multiplets and hypermultiplets are factorized under $\mathcal{N}=2$ supersymmetry, the K${\rm \ddot{a}}$hler metric on the full moduli spaces will also appear as a direct product. Then, the K${\rm \ddot{a}}$hler potentials are factorized up to K${\rm \ddot{a}}$hler transformations. Furthermore, the dilaton is a scalar component of a vector multiplet also in heterotic string theory on  $T^{2} \times K3$. As anticipated, we concentrate on the moduli coming from the vector multiplets. To begin with, we ignore the Wilson line moduli, which dualize in type I$^\prime$ to D7 positions. Then, there are only three vector multiplets and their scalar components are 
\begin{eqnarray}
S_{H}&=&B^d + i e^{-2\phi_{10}^{H}}{\rm Vol}(T^{2}\times K3)^{H},\label{eq:het1}\\
T_{H}&=&\int_{T^2}B_{45} + i {\rm Vol}(T^{2})^{H},\label{eq:het2}\\
U_{H}&=&\frac{G_{45}^{H}+i\sqrt{G_{T^{2}}^{H}}}{G_{44}^H}\,,\label{eq:het3}
\end{eqnarray}
where $B^d$ is the axion dual in 4d to $B_{\mu\nu}$ and $4,5$ are the $T^{2}$ directions. \eqref{eq:het1} and \eqref{eq:het2} can be interpreted respectively as the classical action for a 5-brane instanton wrapping the whole internal manifold and a worldsheet instanton wrapping $T^2$. By applying the relations \eqref{eq:Het-I1}, \eqref{eq:Het-I2} to (\ref{eq:het1} -- \ref{eq:het3}), we can obtain the corresponding moduli fields in type I string theory. Because of the Weyl transformation \eqref{eq:Het-I2}, the d-dimensional volume ${\rm Vol}_{d}$ also gets transformed as 
\begin{equation} 
{\rm Vol}_{d}^{H} = e^{-\frac{d\phi_{10}^{I}}{2}}{\rm Vol}_{d}^{I}.
\end{equation}
Hence we have:
\begin{eqnarray}
S_{H}&\longrightarrow&C^d + ie^{-\phi_{10}^{I}}{\rm Vol}(T^{2}\times K3)^{I} =: S_{I},\label{eq:I1}\\
T_{H}&\longrightarrow&\int_{T^2}C_{45} + i e^{-\phi_{10}^{I}}{\rm Vol}(T^{2})^{I} =:S_{I}^{\prime},\label{eq:I2}\\
U_{H}&\longrightarrow&U_{I}\label{eq:I3},
\end{eqnarray}
where $C^d$ denotes the axion dual in 4d to the RR two-form
$C_{\mu\nu}$ and $C_{45}$ denotes the latter form 
on $T^{2}$. $S_{I},S_{I}^{\prime},U_{I}$ are scalar components of the
vector multiplets in type I string theory.\footnote{\cite{Berg:2005ja}
  discussed the one-loop corrections to K${\rm \ddot{a}}$hler
  potentials in terms of the moduli $S_{I},S_{I}^{\prime},U_{I}$ plus
  Wilson line moduli coming from the reduction on $T^{2}$. These results were later generalized in \cite{Haack:2008yb} to include both types of open string moduli of type I$^\prime$ (i.e. positions in $T^2$ of D7 and D3).} Again it
is clear that \eqref{eq:I1} and \eqref{eq:I2} are respectively the
classical action for a D5 instanton wrapping the whole internal
manifold and a D1 instanton wrapping $T^2$.

Let us  move on to the next step, namely the duality between type I string theory and type I$^{\prime}$ string theory. Our ultimate goal is $\alpha^{\prime}$ corrections in F-theory compactifications in the presence of 7-branes. In order to achieve this situation, one can take two T-dualities on $T^{2}$. In doing so, one converts the 16 D9 branes (plus images) and the O9-plane into 16 D7-branes (plus images) and 4 O7-planes in type IIB respectively, the latter being placed in the 4 fixed points of the $\mathbb{Z}_2$ action on the torus. The duality transformations are: 
\begin{eqnarray}
{\rm Vol}(T^{2})^{I} &=& \frac{1}{{\rm Vol}(T^{2}/\mathbb{Z}_{2})^{I^{\prime}}}\,,\label{eq:IIprime1}\\
e^{-2\phi_{10}^{I}} ({\rm Vol}(T^{2})^{I})^{2} &=& e^{-2\phi_{10}^{I^{\prime}}}\,.\label{eq:IIprime2}
\end{eqnarray}
The last equality comes from the requirement that the four dimensional dilaton becomes the same on both sides \cite{Polchinski:1996fm}
\begin{equation}
e^{-2\phi_{10}^{I}}{\rm Vol}(T^{2}\times K3)^{I} = e^{-2\phi_{10}^{I^{\prime}}}{\rm Vol}(T^{2}/\mathbb{Z}_{2}\times K3)^{I^{\prime}}. \label{eq:T-dual}
\end{equation}
After rewriting (\ref{eq:I1} -- \ref{eq:I3}) in terms of the variables in type I$^{\prime}$ string theory through the relations \eqref{eq:IIprime1} and \eqref{eq:IIprime2}, we have the relations between moduli on both sides:
\begin{eqnarray}
 S_{I}&\longrightarrow&\int_{K3}C_4+ie^{-\phi_{10}^{I^{\prime}}}{\rm Vol}(K3)^{I^{\prime}}=:T_{I^{\prime}}\,,\label{KaehlMod}\\
 S_{I}^{\prime} &\longrightarrow& C_0+ie^{-\phi_{10}^{I^{\prime}}} =:S_{I^{\prime}}\,,\label{AxioDil}\\
 U_{I} &\longrightarrow& U_{I^{\prime}}.
\end{eqnarray}
One readily sees that \eqref{KaehlMod} and \eqref{AxioDil} are the classical actions of an Euclidean D3-brane wrapping $K3$ and of a D(-1) instanton respectively. These are the three moduli we are most interested in: The first is the standard complexified K\"ahler modulus for $K3$, whose imaginary part is of order $g_s^{-1}\alpha^{\prime -2}$; the second is the usual axio-dilaton, whose imaginary part is $g_s^{-1}$; the third is the complex structure modulus of $T^2$.

Let us now consider Wilson line moduli in heterotic string theory. We take them to be defined as
\be\label{WilsonLinesH}
A^i_H:=U_H A^i_4-A^i_5\qquad\qquad i=1,\ldots,16\,,
\ee
where $A^i_{4,5}$ are the components of the i-th vector in the Cartan torus of the heterotic gauge group $SO(32)$ along the directions of $T^2$. They trivially map under S-duality to Wilson line moduli $C^i_{I} $along $T^2$ of the vector fields living on the 16 D9-branes of type I. The latter, in turn, map under the two T-dualities to the positions of the 16 D7-branes of type I$^\prime$ on $T^2/\mathbb{Z}_2$:
\bea\label{WilsonLinesIIB}
C^i_{I}\longrightarrow U_{I^\prime}p^i_4-p^i_5\,.
\eea

\section{Threshold corrections and \s invariance}\label{ThresCorr}

Let us now systematically analyze the threshold corrections to the K\"ahler metric of the vector multiplet moduli space of type I$^\prime$ string theory. As anticipated, the $\cN=2$ supersymmetry allows us to extract all these corrections from the ones of the prepotential. The K\"ahler potential expressed in terms of $\cF$ is:
\bea\label{KaehlerPotPrep}
K&=&-\log\,i\left[2\cF-2\bar\cF-\sum_{\alpha}(\phi^\alpha-\bar\phi^\alpha)\left(\partial_{\phi^\alpha}\cF+\partial_{\bar\phi^\alpha}\bar\cF\right)\right]\,,
\eea
where $\phi^a$ are all the scalars of the vector multiplets. We will therefore use  known results for the corrections to the prepotential in heterotic string theory and translate them to corrections to the K\"ahler potential of type I$^\prime$ using the duality dictionary of subsection \ref{DualityDictionary}. Moreover, we will analyze the \s properties of the results, showing invariance for the K\"ahler potential up to K\"ahler transformations.

In the orbifold\footnote{Which orbifold and how the orbifold action is embedded in the gauge degrees of freedom are all information affecting the low energy physics in the hypermultiplet sector, and they do not enter the prepotential for vector moduli \cite{Harvey:1995fq}, which we are interested in here.} limit of K3, CFT techniques have been used in the heterotic side to compute explicitly all \a corrections, perturbative and non-perturbative \cite{Harvey:1995fq, Henningson:1996jz}. However, the orbifold limit is incompatible with the large volume expansion, as some 2-cycles of $K3$ are shrinking to zero size. Nevertheless, this will not affect our type IIB analysis because, as mentioned, those 2-cycles produce moduli in the hypermultiplet moduli space. Therefore \a corrections to the latter become important, but \a corrections to the vector multiplet moduli space are still subdominant in the orbifold limit of K3.

\subsection{Ignoring Wilson lines}
 
We begin by analyzing the easier case in which we consider the region
of the moduli space where all the Wilson line moduli of heterotic
string theory are vanishing. We will consider a type I$^{\prime}$
string theory dual to a particular type of the
Bianchi-Sagnotti-Gimon-Polchinski model \cite{Bianchi:1990yu,
  Gimon:1996rq}. We may maximally have an $SU(16)$ gauge group in a
special region we have chosen of the hypermultiplet moduli
space.\footnote{This is due to the lack of vector structure arising
  from the particular embedding in the gauge degrees of freedom of the
  orbifold action which describes the K3
  (see~appendix~\ref{app:prepotentials}).}
The $C^i$ moduli introduced above will locally parameterize the
directions normal to the $SU(16)$ region in the moduli space of type
I$^\prime$ theory. In our model, the tadpole cancellation condition is
satisfied without including \emph{mobile} D3-branes. Therefore, we do
not have D3-brane moduli in our type I$^{\prime}$ string theory simply
because all D3-branes needed for tadpole cancellation are stuck at the
16 orbifold points of $T^4/\mathbb{Z}_2$ and have no deformation
moduli along the $T^2$ either (see
appendix~\ref{app:prepotentials}).\footnote{In the dual heterotic
  string theory, we have 16 small instantons
  \cite{Witten:1995gx,Aldazabal:1997wi}. They dualize to 16 rigid,
  space-filling half-D3-branes with total charge of 8.}

In the heterotic model at hand, the prepotential $\cF$ has been computed to all orders in \a using CFT techniques in \cite{Harvey:1995fq, Henningson:1996jz}. Due to the holomorphicity of the prepotential and to the fact that the real axionic shift $S_H\to S_H+\lambda$ is an exact symmetry of the perturbative theory, $\cF$ is exact already at one-loop order in perturbation theory for the heterotic string coupling constant contained in $1/\I S_H$. Thus, up to non-perturbative corrections in $1/\I S_H$, the result is:
\begin{align}\label{PrepHet}
  \begin{split}
    \cF_H(S_H,T_H,U_H)&=\hat S_HT_HU_H+h(T_H,U_H)\, ,\\
    \hat S_H&=S_H+\cfrac12\partial_{T_H}\partial_{U_H}h(T_H,U_H)\,,
  \end{split}
\end{align}
where $\hat S_H$ is the corrected $S_H$ modulus, at all orders in \a
\cite{deWit:1995zg} (see also
\cite{Antoniadis:1995jv}).\footnote{Actually, as explained in
  \cite{deWit:1995zg}, one has to further require the difference
  $S_H-\hat S_H$ to be finite throughout the $(T_H,U_H)$ moduli space,
  in order that the value of $S_H$ still plays the role of the
  universal string-loop counting parameter. This condition leads to
  the addition of a counterterm in the definition of $\hat S_H$,
  which, being modular invariant, will not be important for our
  analysis.} Like for the prepotential, $S_H$ is corrected only at
one-loop in string perturbation theory.

Before giving the definition of the function $h$, we can directly write the prepotential in type I$^\prime$ string theory using the dictionary given in subsection \ref{DualityDictionary}. One caveat must be made, though\footnote{We thank James Gray and Ioannis Florakis for pointing this issue out and for related discussions.}. We are going to assume that this dictionary does not itself receive quantum corrections, at least in perturbation theory for $1/\I S_H$. The fact that the result we find via duality exactly contains the corrections found in \cite{Berg:2005ja} via a genuine type I computation suggests that at least the heterotic/type I S-duality is robust against quantum corrections. Moreover, since we make two T-dualities along a \emph{factorized} $T^2$, makes us confident that also the T-duality step is safe. Thus we have:
\begin{align}\label{PrepIIB}
  \begin{split}
    \cF(S,\hat T,U)&=S\hat TU+h(S,U)\,,\\
    \hat T&=T+\cfrac12\partial_{S}\partial_{U}h(S,U)\,.
  \end{split}
\end{align}
Notice that,  as a consequence of the exactness (both perturbatively and non-perturbatively) of \eqref{PrepHet} in $1/\I T_H$ (i.e. in \a), the corresponding type IIB expression above is exact in $g_s=1/\I S$. Namely it contains all perturbative and non-perturbative corrections in the type IIB string coupling constant. However, since \a is only contained in $T$, \eqref{PrepIIB} does not contain non-perturbative \a corrections, because \eqref{PrepHet} is up to non-perturbative corrections in $1/\I S_H$. $\hat T$ is analogously the corrected type IIB K\"ahler modulus at all order in \g, but only perturbatively in \a (one-loop is again the only non-trivial contribution). 

The function $h$ has a very explicit expression in terms of tri-logarithmic functions. The one valid in the region $\I S> \I U$ is (see appendix \ref{app:UniversalPrepotential} for the computation):
\be\label{hFunction}
h(S,U)=-\frac{i}{(2\pi)^4}\left[\Li_3\left(e^{2\pi i(S-U)}\right)+\sum_{\substack{k,l\ge0 \\ (k,l)\neq(0,0)}} c(kl)\Li_3\left(e^{2\pi i(kS+lU)}\right)\right]+\frac{15i}{2\pi^4}\zeta(3)+\frac{U^3}{12\pi}\,,
\ee
where 
\bea
\Li_m(z)=\sum_{n=1}^{\infty}\frac{z^n}{n^m}\,,\qquad\qquad\sum_{n=-1}^{\infty}c(n)z^n=\frac{E_6E_4}{\eta^{24}}(z)\,,
\eea
$E_{6,4}$ and $\eta$ being the usual Eisenstein series and Dedekind function respectively. In order to extend $h$ to the complement of the $S,U$ moduli space, one performs an analytic continuation. This leads simply to the expression $h(U,S)$ (i.e. \eqref{hFunction} with $S$ and $U$ exchanged) valid in the region $\I U> \I S$. The two expressions clearly connect at the branch locus $S=U$. Let us remark that the expression \eqref{KaehlerPotPrep} for the K\"ahler potential in terms of the prepotential is invariant  under shift of $\cF$ by any polynomial at most quadratic in the $\phi^a$, with real coefficients. As a consequence, the functions $h(S,U)$ and $h(U,S)$ are defined up to a polynomial at most quadratic in $S,U$ with real coefficients. This ambiguity is related to non-trivial quantum monodromies. In special regions of the $(S,U)$ moduli space the function $h$ develops logarithmic singularities. This is due to the fact that some massive vector multiplets which have been integrated out become massless on these loci and thus have to be included among the low energy excitations. Correspondingly, the gauge group gets enhanced. In particular, from $U(1)\times U(1)$ corresponding to the $S,U$ moduli, one has $SU(2)\times U(1)$ along the codimension one locus $S=U$ and $SO(4)\,,\,SU(3)$ on the codimension two loci $S=U=i\,,\,S=U=\rho\,(=e^{2\pi i/3})$ modulo \s respectively. This phenomenon results in a modification of the classical duality group due to non-trivial monodromies around the singular loci \cite{Antoniadis:1995jv}. The duality group must not change the physical metric. This means that the prepotential will generically transform covariantly under the duality group up to a shift by polynomials at most quadratic in $S,U$ with real coefficients. The specific form of these monodromies will not be of interests to us and thus in the sequel we will focus only on the modular properties of the prepotential.

Using the quantum corrected $\cN=2$ K\"ahler variables $S, \hat{T},U$, we can now insert \eqref{PrepIIB} in \eqref{KaehlerPotPrep} and expand the logarithm. We thus obtain the full quantum K\"ahler potential of type I$^\prime$ theory, up to non-perturbative \a corrections:
\bea\label{KaehlerPotCorrections}
K(S,T,U)&=&K^{(0)}(S,T,U)+\sum_{n=1}^{\infty}\frac{1}{n}K^{(n)}(S,T,U)\,,\nn\\
K^{(0)}(S,T,U)&=&-\log\left[-i(S-\bar S)(T-\bar T)(U-\bar U)\right]\,,\nn\\
K^{(n)}(S,T,U)&=&-\frac{(-1)^n}{(T-\bar T)^n}\left[\frac{2h-2\bar h}{(S-\bar S)(U-\bar U)}-\frac{\partial_Sh+\partial_{\bar S}\bar h}{U-\bar U}-\frac{\partial_Uh+\partial_{\bar U}\bar h}{S-\bar S}\right.\nn\\ &&\qquad\qquad\quad\left.-\cfrac12(\partial_S\partial_Uh-\partial_{\bar S} \partial_{\bar U}\bar h)\right]^n\,.
\eea
For reasons that will be clear shortly, in this expression we kept the dependence on $T$, even though at the quantum level ($n>0$) the latter is not anymore a good K\"ahler variable and it must be replaced by $\hat T$. Of course in  \eqref{KaehlerPotCorrections} one has to pick the right convergent expression for the function $h$, depending on which region of the $(S,U)$ moduli space one is looking at. We can appreciate the easy structure of such corrections. First of all, the \a parameter is only appearing in the classical $T$ modulus in front, and only \emph{even} powers of \a are present (because $1/\I T$ is of order \a$^2$). Hence the famous \a$^3$ correction computed in \cite{Becker:2002nn} is not included in \eqref{KaehlerPotCorrections}.\footnote{To form odd powers of \a one would need to use the K\"ahler modulus for the torus $T^2$, which in our case belongs to the hypermultiplet moduli space.} This is explained by the fact that this  \a$^3$ correction is proportional to the Euler characteristic of the type IIB Calabi-Yau threefold, which in our case is vanishing, because the threefold is $K3\times T^2$. Another important feature is that, at the perturbative level for the string coupling constant, only \emph{even} powers of \g appear in \eqref{KaehlerPotCorrections}. This is due to the fact that the function $\partial_Sh$ goes to zero in the perturbative limit for $1/\I S$:
\be
\partial_Sh\longrightarrow0\quad{\rm exponentially\; \;for}\quad S\longrightarrow i\infty\,,
\ee
where we used the following property of logarithmic functions:
\be
\frac{\rm d}{{\rm d}z}\Li_m(z)=\frac{1}{z}\Li_{m-1}(z)\,.
\ee
Therefore in $K^{(n)}$ only terms which have an overall factor $\left(\frac{1}{\I T\I S}\right)^n$ in front survive, which means, recalling definitions \eqref{KaehlMod} and \eqref{AxioDil}, two powers of \g. This is explained by the fact that we are freezing open string moduli, thus neglecting the effect of 7-branes on the bulk low energy fields. The latter indeed induces also odd powers of \g and we will take them into account in the next subsection (see \eqref{FTheoryTreealpha} for the F-theory picture at tree level in \a). 

As one immediately sees the K\"ahler modulus and the axio-dilaton, while decoupled at tree level in \a, already mix at the first non-trivial \a order. In particular, one can recognize in \eqref{KaehlerPotCorrections} the threshold correction of \cite{Berg:2005ja} at \a$^2$ order ($n=1$) (see also \cite{Berg:2007wt}):
\be\label{BHKcorrection}
K^{(1)}=-\frac{\cE(U,\bar U)}{(T-\bar T)(S-\bar S)}\qquad\qquad\cE:=\lim_{S\to i\infty}\frac{2h-2\bar h}{U-\bar U}-\partial_Uh-\partial_{\bar U}\bar h\,.
\ee
This perturbative correction comes from the joint contribution of two different kinds of BPS states: The Kaluza-Klein states exchanged between the D7-branes and the non-mobile D3-branes (also viewable as one-loop of open strings) and the non-orientable open stings with M\"obius strip topology stretched between parallel D7-branes. Notice that, in contrast to ref. \cite{Berg:2007wt}, in \eqref{KaehlerPotCorrections} there is no correction proportional to $\left(\frac{1}{\I T}\right)^2$ with no power of $\I S$. This is because the latter correction would come from the exchange of strings wound along circles in the intersection of two D7-branes; But those circles are not there in our situation, because D7's are just points on $T^2$, thus they either do not intersect or they coincide, and $K3$ has no non-trivial 1-cycles.

By looking at \eqref{KaehlerPotCorrections} one can easily infer which
kind of correction occupies a given point in
fig.~\ref{fig:corrections}. For instance, there is no non-perturbative
\g correction in the first \a tower, namely tree level in \a. This
property still holds after the inclusion of Wilson line moduli. With
no Wilson line moduli, only the lowest \g order is non-zero in the
first \a tower. Wilson line moduli will only add perturbative \g
corrections.  Non-perturbative \g corrections are instead present at
all non-trivial orders in \a. However, again because of the absence of
Wilson line moduli, there is just one perturbative \g correction for
each \a tower (i.e. for each value of the integer $n$): For the
relative order \a$^{2n}$, such a correction is of relative\footnote{Of
  course we mean relative with respect to the first \a
  tower. Analogously \g powers are relative to the string tree level
  power.} order $g_s^{2n}$.

As a final comment, let us stress again that
\eqref{KaehlerPotCorrections} does not include non-perturbative
corrections in \a. However, worldsheet and D1 instantons are not
present, because they are projected out by the
orientifold.\footnote{More precisely, F1 and D1 with one leg along the
  $T^2$ in principle survive the projection, but there is no
  non-trivial circle in K3 to wrap the other leg around.} On the
other hand, corrections from the \s-invariant euclidean D3 instantons
wrapped on $K3$ are missing in \eqref{KaehlerPotCorrections} and will
be briefly discussed in section \ref{NonPerturbAlpha}. Euclidean D3
branes wrapping $T^2$ times a 2-cycle of K3 correct the metric of the
hypermultiplet moduli space and will not be discussed here. D(-1)
instanton corrections, instead, which are non-perturbative only in \g,
are contained in \eqref{KaehlerPotCorrections}.

\subsubsection*{\s invariance}

Let us now analyze the \s properties of  \eqref{KaehlerPotCorrections}. First of all we notice that $K$ is perfectly symmetric under exchange of $S$ and $U$, at each \a order, thanks to the symmetry of the function $h$ (taking into account that this $\mathbb{Z}_2$ symmetry also changes the region of convergence of $\Li_3$). Therefore an \s invariance for $S$ would automatically imply an \s invariance for $U$. Let us then verify that this invariance is actually there, at each \a order. To see this, it is enough to check invariance for the two generators of \s, namely
\be
{\rm T}=\left(\begin{array}{cc}1&1\\0&1\end{array}\right)\,,\qquad\qquad {\rm S}=\left(\begin{array}{rr}0&-1\\1&0\end{array}\right)\,.
\ee
Under T-transformations $h$ is invariant in either region of the $S,U$ moduli space ($h(S,U)$ is obviously invariant and $h(U,S)$ is invariant up to a quadratic polynomial in $S$ with real coefficients, which, as said, is immaterial for the K\"ahler potential). Under S-transformations, $h$ transforms as follows \cite{Harvey:1995fq}:
\be\label{hSTransformation}
h\overset{\rm S}{\longrightarrow}\frac{h}{S^2}\,,
\ee
where on both sides one has to use the right definition of $h$ (for instance, if one begins in the region $\I S>\I U$ and the S-transformation sends to the other region, one has $h(U,-1/S)=h(S,U)/S^2$). On the other hand, the classical $T$ modulus is invariant under \s. Indeed, $\I T$ is $e^{-\phi}$ times the volume of $K3$ in the \emph{string frame}, that means that it is simply the volume of $K3$ in the Einstein frame, which is \s invariant (as the Einstein frame metric is \s invariant). It is then easy to check, using \eqref{hSTransformation}, that $K^{(0)}$ and each of the $K^{(n)}$ in \eqref{KaehlerPotCorrections} are separately \s invariant. In summary at each perturbative \a order the K\"ahler potential of type I$^\prime$ string theory is invariant under the following group:
\bea\label{SUDualityGroup}
O(2,2,\mathbb{Z})=SL(2,\mathbb{Z})_S\times SL(2,\mathbb{Z})_U\rtimes\mathbb{Z}_2\,.
\eea

\subsection{Inclusion of Wilson lines}

Let us now include the $16$ Wilson line moduli $C^i$ defined in \eqref{WilsonLinesIIB}. This will generically break the gauge group to $U(1)^{15} \subset SU(16)$. In total,
the vector multiplet moduli space will have $19$ complex dimensions. Before introducing the K\"ahler potential, we must say that in the presence of Wilson line moduli the axiodilaton $S$ is no longer a good $\cN=2$ K\"ahler coordinate, but it has to be replaced by \cite{Berg:2005ja} 
\be\label{HatS}
\hat S=S+\frac{1}{2}\sum_{i=1}^{s}C^i\frac{C^i-\bar C^i}{U-\bar U}\,,
\ee
where $s=0,\ldots,16$ indicates the number of Wilson lines we have turned on.
The exact prepotential up to non-perturbative \a corrections looks like
\be\label{QuantumPrepWilson}
\cF=\hat S\hat TU-\frac{\hat T}{2}\sum_{i=1}^{s}(C^i)^2+\tilde h(\hat S,U,C^i)\,, 
\ee
where the $\tilde h$ function is given by \eqref{general} with a particular embedding of the orbifold action, with the arguments $(\bar{y}, y^+, y^-)$ transformed into $(C^i, \hat{S}, U)$ respectively and also with an appropriate normalization (see appendix \ref{app:SpecificWL} for explicit formulae). Moreover, the quantum corrected $T$-modulus 
has the following general expression \cite{Harvey:1995fq}
\bea\label{QuantumTGeneral}
\hat T&=&T+\frac{1}{s+4}\left[2\partial_{\hat{S}}\partial_{U}\tilde{h}(\hat{S},U,C^i)-\sum_{i=1}^{s}\partial_{C^i}\partial_{C^i}\tilde{h}(\hat{S},U,C^i)\right]\,,
\eea
Equation \eqref{QuantumTGeneral} reduces, as it should, to the second equation of \eqref{PrepIIB} for $s=0$ (absence of Wilson line moduli), and has to be used with $s=16$ in the most generic case.

Note that the obtained prepotential \eqref{QuantumPrepWilson} is the generalization of \cite{Berg:2005ja} including the non-perturbative terms in $g_s$ if one uses the orbifold action \eqref{gamma} for $T^4/\mathbb{Z}_2$ embedded in a sixteen dimensional self-dual lattice of the gauge degrees of freedom. The explicit expression of the prepotential is \eqref{explicit1} without Wilson line moduli and \eqref{pre.wil} with Wilson line moduli, with $(\hat{T}_H, U_H)$ transformed into $(\hat{S}^{\prime}_I, U_I)$. By taking a weak coupling limit $(\hat{S}^{\prime}_I)_2 \rightarrow \infty$, one can show that the prepotentials \eqref{explicit1} and \eqref{pre.wil} exactly reduce to the ones obtained in \cite{Berg:2005ja}. Since the explicit computation is rather technical and not relevant here, we will postpone it to appendix \ref{app:SpecificWL}. 

Let us now focus on the first \a tower of fig. \ref{fig:corrections}. Again the volume modulus $T$ decouples, as it factorizes in the prepotential
\be\label{TreeLevelPrepotential}
\cF_{\rm tree}=T\left(\hat SU-\frac{1}{2}\sum_{i=1}^{16}(C^i)^2\right)\,.
\ee
Using eq. \eqref{KaehlerPotPrep}, the K\"ahler potential at tree level in \a looks like
\bea\label{TreeLevelKaehlerWilson}
K^{(0)}&=&-\log[-i(T-\bar T)]-\log\left[(\hat S-\bar{\hat S})(U-\bar U)-\frac{1}{2}\sum_{i=1}^{16}(C^i-\bar C^i)^2\right]\nn\\ &=&
-\log[-i(T-\bar T)]-\log\left[(S-\bar{S})(U-\bar U)\right]\,.
\eea
As it is shown in the last equality above, this K\"ahler potential is still invariant under the duality group \eqref{SUDualityGroup} (up to a K\"ahler transformation). While the second line in eq. \eqref{TreeLevelKaehlerWilson} makes manifest the modular properties of $K^{(0)}$, one has to use the first line to compute the K\"ahler metric because in this case, as said, $S$ is not a good $\cN=2$ special coordinate any more, whereas $\hat S$ is. In fact, the duality group in the presence of Wilson line moduli gets enlarged from \eqref{SUDualityGroup} to $O(2,2+s,\mathbb{Z})$. By embedding $SO(2,2,\mathbb{Z})$ into $SO(2,2+s,\mathbb{Z})$, one realizes \cite{deWit:1995zg} that the duality group which was rotating only the axiodilaton in the absence of Wilson lines, i.e. $SL(2,\mathbb{Z})_S$, generalizes to the following group of transformations acting on $\hat{S}$ and touching the $U$ and the $C^i$ moduli as well
\bea\label{SL2ZWL}
\hat{S}&\longrightarrow&\frac{a\hat{S}+b}{c\hat{S}+d}\nonumber\,,\\
U&\longrightarrow&U-\frac{c}{2(c\hat{S}+d)}\sum_{i=1}^sC^iC^i\nonumber\,,\\
C^i&\longrightarrow&\frac{C^i}{c\hat{S}+d}\nonumber\,,\\
\eea
where $a,b,c,d$ are the integral entries of an $SL{(2,\mathbb{Z})}$ matrix. It is easy to see that the group of transformations \eqref{SL2ZWL} leaves invariant $K^{(0)}$ expressed in terms of the good $\cN=2$ coordinates (first line of \eqref{TreeLevelKaehlerWilson}). 
Another important property of $K^{(0)}$ is that still it does not undergo non-perturbative corrections in the type IIB string coupling \g$=1/{\rm Im} S$. On the other hand, the presence of Wilson line moduli seems to introduce perturbative corrections in \g. Indeed, already at tree level in \a, we can perform the expansion
\bea\label{ExpansionWilson0}
K^{(0)}=-\log\left[-i(T-\bar T)(\hat S-\bar{\hat S})(U-\bar U)\right]-\sum_{k=1}^{\infty}\frac{(-1)^k}{k}\left(\frac{\sum_{i}(C^i-\bar C^i)^2}{2(\hat S-\bar{\hat S})(U-\bar U)}\right)^k\,,
\eea
where string loop corrections are explicit in
the last term. However, the second line in \eqref{TreeLevelKaehlerWilson} shows that, at least at the level of the K\"ahler potential, such corrections can be reabsorbed in the old definition of the axiodilaton, thus implying that also in this case the K\"ahler potential at tree level in \a is exact in \g.

Eq. \eqref{ExpansionWilson0} is the analog of the general expansion \eqref{FTheoryTreealpha} ensuing from the F-theory Calabi-Yau fourfold. However, as we will see explicitly in section \ref{FTheoryPicture}, the particular model at hand lifts to F-theory on an elliptic K3, whose period structure is far easier than the fourfold one. For this reason, \eqref{ExpansionWilson0} is an exact expression in \g  and no Sen's weak coupling limit is required to write it. 

Analogously, for higher \a towers, $K^{(n)}$ will not
contain only one perturbative \g correction, but many others again due
to 7-brane deformations which are now included in the calculation (in
the previous section these degrees of freedom were frozen by the
condition $C^i=0$). Indeed, inserting \eqref{QuantumPrepWilson} in
\eqref{KaehlerPotPrep} and expanding the logarithm one finds
\be\label{KnWilsonLine}
K^{(n)}=-\sum_{k=n}^{\infty}\frac{n(-1)^{k}}{k(T-\bar T)^n}\binom{k}{n}\frac{\sum_{i}(C^i-\bar C^i)^{2(k-n)}\mathbf{h}^n}{2^{(k-n)}(\hat S-\bar{\hat S})^k(U-\bar U)^k}\,,
\ee
where
\be
\mathbf{h}=2\tilde h-2\bar {\tilde h}-\sum_{\phi=\hat S,U,C^i}(\phi-\bar \phi)(\partial_{\phi}\tilde h+\partial_{\bar{\phi}}\bar {\tilde h})-(\hat{T}-\bar{\hat{T}}-T+\bar{T})[(\hat{S}-\bar{\hat{S}})(U-\bar{U})-\cfrac12\sum_{i}(C^i-\bar C^i)^2]\,.
\ee
Eq. \eqref{KnWilsonLine} reduces to the last of eq. \eqref{KaehlerPotCorrections} by putting $C^i=0$, because only the term $k=n$ of the sum survives. In the presence of Wilson lines all the following terms of the sum seem to contain an infinite amount of perturbative \g corrections. However, again, we may try to get rid of them, at least at the level of the K\"ahler potential, by rewriting them in terms of the old axiodilaton $S$. In this way, the K\"ahler potential looks exactly like the one in \eqref{KaehlerPotCorrections}, with
\bea
K^{(n)}(S,T,U,C^i)&=&-\frac{(-1)^n}{(T-\bar T)^n}\left[\frac{\mathbf{h}(\hat S(S),U,C^i)}{(S-\bar S)(U-\bar U)}\right]^n\,.
\eea
In the specific example with Wilson lines presented in appendix \ref{app:SpecificWL}, we will indeed see that the dependence on $S$ of the function ${\bf h}$ above does not introduce any further perturbative \g corrections in each \a tower. We believe that the same conclusion holds more generally.

\subsubsection*{\s invariance}

We have already seen the \s invariance of the K\"ahler potential at tree level in \a, i.e. $K^{(0)}$. To prove the invariance for each of the higher \a towers one needs to work out the modular properties of the function $\tilde h(\hat{S},U,C^i)$ under the \s duality \eqref{SL2ZWL}. Luckily, a quick argument allows us to avoid any hard computation. As mentioned, in the presence of Wilson lines the target space duality of the dual heterotic string theory enhances to $O(2,2+s,\mathbb{Z})$. This is an exact symmetry of the effective action at all orders in perturbation theory \cite{Sen:1994fa}. This means that the full K\"ahler potential is invariant under this group (and in particular under its \s subgroup \eqref{SL2ZWL}) up to K\"ahler transformations. To see that the invariance actually holds separately for each \a tower, we just have to remember that the various towers are labeled by different powers of the $T$-modulus. The latter, in turn, is left invariant by $O(2,2+s,\mathbb{Z})$, being the dual to the heterotic axiodilaton $S_H$. In other words, target space dualities do not mix $S_H$ with the other moduli. This concludes the argument and shows \s invariance for each \a tower independently, even in the presence of Wilson line moduli.

\subsection{Non-perturbative \a corrections}\label{NonPerturbAlpha}

The last set of corrections to the vector multiplet metric of type I$^\prime$  string theory which we have not yet discussed are the ones coming from euclidean D3-brane wrapping K3. They are non-perturbative in both \a and \g as they involve the exponential of the $T$ modulus and they must be trivially \s-invariant, as their sources are singlets.

The exact prepotential for the type I$^\prime$ model is \cite{Berglund:2005dm}
\be
\cF=\hat S\hat TU-\frac{\hat T}{2}\sum_{i=1}^{16}(C^i)^2+\tilde h(\hat S,U,C^i)+\sum_m\cA_m(\hat S,U,C^i)e^{2\pi i m T}\,, 
\ee
where $m$ is the instanton charge. The $\cA$ factors may for example be computed using the duality of type I$^\prime$ theory with type IIA on a Calabi-Yau threefold which admits a K3 fibration over a two-sphere. A partial computation of these terms from this perspective was provided in \cite{Berglund:2005dm}. This duality originates from the six-dimensional one between heterotic on $T^4$ and type IIA on K3 \cite{Harvey:1995rn,Kachru:1995wm,Ferrara:1995yx}, just by fiberwise iteration on an $S^2$. Under this duality, D3-branes wrapping K3 on the type I$^\prime$ side are mapped to world-sheet instantons wrapping the base $S^2$ on the type IIA side.

\section{F-theory picture}\label{FTheoryPicture}

\subsection{Preliminaries}

In this section we will describe the F-theory counterpart of the type
IIB picture given in section \ref{ThresCorr}, by making use of the
F/M-theory duality. Before we begin, a comment concerning the F-theory
limit of M-theory compactification on elliptically fibered Calabi-Yau
manifolds is in order.  To obtain F-theory from M-theory
\cite{Denef:2008wq} one has to send the volume of the elliptic fiber
to zero, which will therefore not be a modulus of the effective theory
of F-theory (see below). Now M2-branes wrapping on the $T^2$ fiber
have tension proportional to $R_{M}R_{T}/l_{M}^3$, where $l_M$ is the
11d Planck length, $R_{M}$ is the radius of the M-theory circle and
$R_{T}$ is the radius of the circle along which we take T-duality to
transform type IIA into type IIB string theory. After the reduction,
those M2-branes become strings wound around the T-duality circle, with
mass proportional to $R_{T}/\alpha^\prime$. On the other hand, they
are mapped by T-duality to KK modes in type IIB string theory with the
same mass. Under the F-theory limit $R_{T}$ goes to $0$. Hence, all
the KK modes become massless and another dimension comes out in the
type IIB side (because the IIB circle has radius
$\alpha^\prime/R_{T}$). From the M-theory perspective, this limit
means that M2-branes wrapped on the vanishing $T^2$ behave as massless
particles and they affect the low energy effective theory as
$l_M$-corrections, i.e. the large volume approximation for the
M-theory compactification clearly breaks down. This important
deviation from 11d supergravity is fully kept by summing up all the KK
modes mentioned above in the type IIB effective theory. The latter
modes are indeed becoming massless as the F-theory limit
decompactifies the third spatial dimension. All the other deviations
are, instead, normally sub-leading as long as all the other volumes of
the elliptic Calabi-Yau are large.

The low energy field theory of the model we have been discussing so far is a four-dimensional $\cN=2$ supergravity. The vector multiplet moduli space of this theory is a $19$-dimensional Special K\"ahler manifold which classically looks like
\be\label{ClassModSpace}
\cM_{\rm V}=\frac{SU(1,1)}{U(1)}\times\frac{O(2,18)}{O(2)\times O(18)\times O(2,18,\mathbb{Z})}\,,
\ee
where the first factor is parameterized locally by the volume modulus $T$. The gravitational multiplet introduces yet another vector field, but no physical scalar is associated to it. Thus we can describe the physical scalar manifold \eqref{ClassModSpace} by embedding it in an ambient $20$-dimensional projective space parameterized by homogeneous coordinates $X^\Lambda$. The theory in this sector is specified by an holomorphic prepotential $F(X)$, which is an homogeneous function of degree two. Moreover one defines the standard symplectic section 
\be\label{SymplSec1}
\mathbb{S}=\left(\begin{array}{c}X^\Lambda\\ \partial_{X^\Lambda} F\end{array}\right)\,,\qquad\qquad\Lambda=\{0,\alpha(=1,\ldots,19)\}\,.
\ee
The manifold $\cM_{\rm V}$ is then taken to be the codimension one  hypersurface with equation
\be
\mathbb{\bar S}\Omega\mathbb{S}=const.\,,\qquad\qquad\Omega=\left(\begin{array}{cc}0&I_{\rm 20x20}\\-I_{\rm 20x20}&0\end{array}\right)\,.
\ee 
In the local patch where $X^0\neq0$, the prepotential can be written as
\be
F(X^\Lambda)=(X^0)^2\cdot\cF(\phi^\alpha)\,,\qquad\qquad\phi^\alpha\equiv\frac{X^\alpha}{X^0}\,,
\ee 
where $\phi^\alpha=\{T,S,U,C^i\}$ is a convenient choice of local coordinates on this hypersurface. Therefore, locally in the moduli space \eqref{ClassModSpace}, we can write the symplectic section \eqref{SymplSec1} as 
\be
\mathbb{S}=X^0\cdot\left(\begin{array}{c}1\\ \phi^\alpha \\ 2\cF -\sum_\alpha\phi^\alpha\partial_{\phi^\alpha}\cF\\ \partial_{\phi^\alpha} \cF\end{array}\right)\,.
\ee
The K\"ahler potential for the vector multiplet moduli space is then of the form
\be
K=-\log\left(i\mathbb{\bar S}\Omega\mathbb{S}\right)\,,
\ee
which, up to a K\"ahler transformation, is equal to \eqref{KaehlerPotPrep}. The discrete reparameterizations $O(2,18,\mathbb{Z})$ in \eqref{ClassModSpace} can be embedded in the group of symplectic rotations of $\mathbb{S}$, i.e. $Sp(40)$, which clearly leaves all the physical quantities invariant. The duality group \eqref{SUDualityGroup} acting on the $S,U$ moduli only is a subgroup of $O(2,18,\mathbb{Z})$.

\subsection{F-theory lifts}\label{FLifts}

Let us now come to F-theory. It is known that the heterotic $SO(32)$ theory compactified on $T^2$ is dual to F-theory on an elliptically fibered K3 which admits another global section except for the zero section \cite{Aspinwall:1996nk}. The model we have been discussing so far is a further K3 compactification of this theory down to four dimensions. The type I$^\prime$ theory of the previous sections is related by two T-dualities to the BSGP model, which in turn is S-dual to an heterotic $SO(32)$ theory \emph{without} vector structure. In the absence of Wilson lines, the maximal surviving gauge group is $SU(16)$ rather than $SO(32)$. As argued below, we may forget about this subtlety when focusing on the vector multiplet moduli space.

It turns out \cite{Berkooz:1996iz} that the heterotic $SO(32)$ theory without vector structure is dual to the  heterotic $E_8\times E_8$ theory with instanton embedding $(12,12)$. At generic points of the hypermultiplet moduli space of the dual pair the non-Abelian gauge group is completely broken by the vevs of the instanton moduli (hypermultiplets) and only an $U(1)^4$ factor is left (which corresponds to the three vector moduli $S_H,T_H,U_H$ and the graviphoton). On the other hand, the heterotic $SO(32)$ theory with vector structure happens to be dual to  the  heterotic $E_8\times E_8$ theory with a \emph{different} kind of instanton embedding. As there will not be  enough instantons for a complete higgsing, non-trivial Wilson lines need to be turned on to break the gauge group to $U(1)$s. For example, in the extreme case of the instanton embedding $(24,0)$, the left gauge theory can be completely higgsed by instantons, while we need $8$ Wilson line moduli which, by taking non-trivial vevs, break the right $E_8$ to the Cartan torus $U(1)^8$. 

We are analyzing in this paper only the vector multiplet moduli space of these theories. In particular, when all Wilson lines are turned-off, the prepotential of the theory without vector structure \eqref{explicit1} perfectly matches the one of the theory with vector structure \eqref{HM}, regardless of them being dual to $E_8\times E_8$ theories with different instanton embeddings and of the consequent fact that we have different gauge groups in four dimensions. This can be explained using the relation between the prepotential for vector multiplets and the supersymmetric index, which, in the absence of Wilson lines, turns out to be insensitive to the instanton embedding \cite{LopesCardoso:1996nc} (see also appendix \ref{app:UniversalPrepotential} for a detailed discussion about this fact).

In this paper, heterotic string theory is compactified on K3$\times T^2$. Consider for simplicity the regime in which the K3 manifold admits an elliptic fibration over $\mathbb{P}^1$.\footnote{The dualities hold throughout the moduli space. The elliptic fibration limit just allows to derive the duality from an adiabatic argument.} The theory admits three possible F-theory duals:
\begin{enumerate}
\item Using the prototype, eight-dimensional duality between heterotic on $T^2$ and F-theory on K3 and upon further compactification on K3, we obtain F-theory on K3$\times$K3. 
\item By applying the 8d duality fiberwise for the heterotic K3 (which we took elliptically fibered) we find a 6d duality with F-theory compactified on $X_3\times T^2$, where $X_3$ is a Calabi-Yau threefold admitting a K3-fibration over $\mathbb{P}^1$ (the $T^2$ is just a spectator here). It turns out that $X_3$ is also elliptically fibered. But most importantly, it is the \emph{same} \cite{Louis:1996mt} Calabi-Yau threefold on which we compactify the type IIA dual to heterotic on an elliptic K3$\times T^2$ \cite{Kachru:1995wm}. The duality with type IIA is described in appendix \ref{DualityHet/IIA}. The base of $X_3$ as an elliptic fibration is an Hirzebruch surface $\mathbb{F}_n$ where $n$ is related to the instanton embedding of the dual $E_8\times E_8$ heterotic theory \cite{Morrison:1996na}. The type IIA geometry is smooth if the corresponding heterotic theory has no non-Abelian unbroken gauge group. In the following we are mostly interested in the two geometries (see \cite{Klemm:1995tj,Hosono:1993qy}): 
\begin{itemize}
\item $X_3=W\mathbb{P}_{1,1,2,8,12}(24)$, which has $h^{1,1}=3$, $h^{2,1}=243$ and thus $\chi=-480$. This is the internal manifold of the type IIA dual to heterotic $E_8\times E_8$ with instanton embedding $(12,12)$. This theory has $3$ vector moduli, corresponding to $S_H,T_H,U_H$ and, in the Higgs phase, has no non-Abelian unbroken gauge symmetries.
\item $X_3=W\mathbb{P}_{1,1,12,28,42}(84)$, which has $h^{1,1}=11$,
  $h^{2,1}=485$ and thus $\chi=-960$. This is the internal manifold of
  the type IIA dual to heterotic $E_8\times E_8$ with instanton
  embedding $(24,0)$. This theory has $11$ vector moduli,
  corresponding to $S_H,T_H,U_H$ as well as the $8$ Wilson line moduli
  needed to completely break the non-Abelian part of the gauge
  symmetry.\footnote{This same theory in the absence of Wilson lines
    would have a non-Abelian gauge group still
    unbroken. Correspondingly the type IIA Calabi-Yau threefold would
    develop singularities, and one has to blow it up before computing
    topological quantities. Since we are only focusing on the vector
    multiplet moduli space, by the argument above we can safely use
    the $X_3$ dual to the $(12,12)$ theory with no Wilson lines, and
    still get the correct answer for the prepotential.}
\end{itemize} 
\item By applying mirror symmetry to type IIA on $X_3$, we get type IIB on the mirror Calabi-Yau $\tilde{X}_3$. The latter theory, which has \emph{no 7-branes}, is equivalent to F-theory on the trivially fibered Calabi-Yau fourfold $\tilde{X}_3\times T^2$. 
\end{enumerate}

Let us discuss the first F-theory lift in light of the quantum corrected prepotential we have in formula \eqref{PrepIIB}. We denote by a prime the F-theory K3 which is elliptically fibered.

\subsection{Classical theory}

We begin by writing the F-theory K\"ahler potential for the vector moduli at tree level in \a. Recalling eq. \eqref{KkKc} and \eqref{KaehlerPot} and that the lift of type I$^\prime$ theory is F-theory on $K3\times K3^\prime$, where $K3^\prime$ is elliptically fibered, one has:
\be\label{KkKc2}
\cK_K=-\log\cV_{K3}-\log\cV_{K3^\prime}\,,\qquad\qquad\cK_c=-\log\int_{K3}\Omega_2\wedge\bar\Omega_2-\log\int_{K3^\prime}\Omega^\prime_2\wedge{\bar\Omega}^\prime_2\,,
\ee
where $\cV$ are the classical volumes in the Einstein frame. As it should be clear from section \ref{ReviewModel}, only the first term in $\cK_K$ and the second in $\cK_c$ of \eqref{KkKc2} enter the K\"ahler potential for vector multiplet moduli. In fact, the vector multiplet moduli are all the moduli of the upper K3 but one ($h^{1,1}-1=19)$. To see this recall that the elliptic fibration defining the upper K3 breaks the ambiguity between its complex and K\"ahler structure. Indeed it selects a particular direction in the space-like three-plane of self-dual harmonic 2-forms in the lattice $\Gamma^{3,19}=H^2(K3^\prime,\mathbb{Z})$ and identifies it with the K\"ahler form, i.e.
\be
J=v^0\omega_0+v\omega\,,
\ee
where $\omega_0$ is the class Poincar\'e dual to the 0-section and $\omega$ is the hyperplane class of the base $\mathbb{P}^1$. This naturally singles out a sublattice $U\subset\Gamma^{3,19}$, spanned by $(\omega_0+\omega,\omega)$, which identifies the K\"ahler moduli of K3$^\prime$. These two classes generate the Picard group, which for a generic K3 is trivial. A choice of a spacelike two-plane in the orthogonal complement $\Gamma^{2,18}$ corresponds instead to a particular complex structure. Thus, the space of the $18$ complex structure deformations of K3$^\prime$ coincides with the second factor in \eqref{ClassModSpace}. Not both of the K\"ahler moduli on  the other hand are physical in F-theory, because the F-theory limit $v^0\to0$ kills one of them. The other one, after normalizing it by the total volume of the internal manifold can be seen to coincide with $T$ of eq. \eqref{KaehlMod} \cite{Grimm:2010ks,Grimm:2011sk}. 

Let us now prove that the K\"ahler potential 
\be\label{K3K3'}
K^{(0)}=-\log\cV_{K3}-\log\int_{K3^\prime}\Omega^\prime_2\wedge{\bar\Omega}^\prime_2
\ee
indeed coincides with \eqref{TreeLevelKaehlerWilson}. The first term of eq. \eqref{K3K3'} clearly coincides, up to a K\"ahler transformation, with the first term in eq. \eqref{TreeLevelKaehlerWilson}. As for the second, let us observe that a convenient parameterization of the periods of K3$^\prime$ can be obtained by applying to $\mathbb{S}$ the symplectic transformation which sends $(T,\partial_T\cF)$ to $(-\partial_T\cF,T)$.  The periods of K3$^\prime$ then coincide with the upper part of the transformed symplectic section. Recalling the expression of the tree level prepotential \eqref{TreeLevelPrepotential}, we thus have\footnote{There is a subtlety in the quantization properties of the chosen basis of $H^2(K3^\prime,\mathbb{Z})$ \cite{Lust:2005bd}. A correctly normalized basis can be found in \cite{Braun:2008ua}.}
\be\label{PeriodsK3'}
\Pi^\Lambda=\left(\begin{array}{c}1\\-SU+\frac{1}{2}\sum_i(C^i)^2\\ S\\ U\\ C^i \end{array}\right)\,.
\ee 
The $C^i$ are identified with the Wilson line moduli, while $U$ is identified with the complex structure of the $T^2$ and $S$ represents the asymptotic axiodilaton. The metric in the chosen basis is of the block-diagonal form
\be
M_{\Lambda\Sigma}={\rm diag}\,\left(\sigma^1,\sigma^1,-I_{\rm 16x16}\right)\,,
\ee
where $\sigma^1$ is the hyperbolic plane metric. It is easy to see now that the second term in eq.  \eqref{K3K3'}, namely $\Pi^\Lambda M_{\Lambda\Sigma}\Pi^\Sigma$, coincides with the second term in eq. \eqref{TreeLevelKaehlerWilson}.

Therefore, classically in \a the K\"ahler modulus is completely decoupled from the complex structure moduli. As already observed, the expression \eqref{K3K3'} is exact in \g thanks to the polynomial structure of the periods of K3$^\prime$ \eqref{PeriodsK3'}. In contrast, in the fourfold case, there is no guarantee that non-perturbative \g corrections are absent in the first \a tower, and thus the perturbative expansion \eqref{FTheoryTreealpha} is only reliable in a regime (the Sen limit) where $\cO(e^{-1/g_s})$-corrections can be consistently neglected.

\subsection{Quantum corrections}

\subsubsection{Sources}\label{Sources}

Let us now turn to quantum corrections, which will destroy the factorization of the K\"ahler modulus and the complex structure moduli in \eqref{K3K3'}, but will still preserve the special K\"ahler geometry of the moduli space. In this respect, it is worth stressing that the classical vector multiplet moduli space \eqref{ClassModSpace} of the theory under consideration exactly matches the classical moduli space of elliptically fibered K3 manifolds. This means that the quantum corrected K\"ahler potential we have in equation \eqref{KaehlerPotCorrections} is interpreted in this F-theory lift as the K\"ahler potential of the \emph{quantum} moduli space of the elliptic K3$^\prime$. Quantum corrections non-trivially mix its complex structure moduli with its unique physical K\"ahler modulus. A natural question is now what are the M-theory BPS objects which generate the corrections we have found in section \ref{ThresCorr}. The answer may be deduced by investigating on the lift of the known BPS objects correcting the K\"ahler metric in type IIB. Such an analysis leads us to the following three sources of corrections:
\begin{enumerate}
\item The non-perturbative \g corrections in
  \eqref{KaehlerPotCorrections} are generated, as said, by D(-1)
  instantons of type IIB string theory. They lift to loops of
  gravitons in eleven-dimensional supergravity. More precisely a D(-1)
  T-dualizes to a D0-brane in type IIA whose worldline wraps the
  T-duality circle of the torus fiber; the latter in turn lifts to a
  KK particle with a unit of momentum along the M-theory circle,
  looping along the T-duality circle of the F-theory fiber.

  In the case of a trivial fibration, these are the types of higher
  derivative corrections to 11d supergravity considered in the series
  of papers
  \cite{Green:1997tv,Green:1997di,Green:1997as,Green:1998by}. In
  particular they contribute to the $R^4$ coupling (four powers of the
  Riemann tensor), which gives the famous $\alpha'^3$ coupling upon
  dimensional reduction \cite{Antoniadis:1997eg, Becker:2002nn}.

  Our results show that the non-trivial fibration structure affects
  this computation in a fundamental way, giving contributions already
  at order $\alpha'^2$.

  Notice that the above corrections are part of those 11d Planck
  length ($l_M$) corrections of the M-theory effective physics which
  stay \emph{finite} in the F-theory limit.\footnote{While these
    corrections certainly represent \a corrections to F-theory
    compactifications, the latter might not all be of this type. See
    for instance \cite{Grimm:2012rg}, where \a effects in the gauge
    coupling function of an F-theory compactification are observed to
    arise from tree-level 11d supergravity, in the presence of
    G-flux.} This limit, indeed, sends $v^0\to0$, where $v^0$ is the
  volume of the elliptic fiber. Given the relation \cite{Denef:2008wq}
  \begin{align}
    \label{lMalphaprime}
    l_M^3=\alpha^\prime\sqrt{v^0}\,,
  \end{align} one deduces that in the quantum corrections to the
  F-theory effective physics the parameter $l_M$ should always appear
  in the finite combination $\alpha'=l^3_M/\sqrt{v^0}$ or powers
  thereof. The loops of 11d supergravitons we are considering here
  should generate corrections scaling in this way in order to survive
  the F-theory limit.  The explicit computation of the corresponding
  quantum correction would go pretty much like the Schwinger loop
  calculation done in \cite{Collinucci:2009nv}, except that in this
  case the elliptic fibration is non-trivial, which as we have seen
  will change the result in important ways. In the next section we
  obtain the final result of this computation via a chain of
  dualities, but it would be interesting to perform the direct
  computation of the graviton loop.

\item The perturbative \g corrections, like the one displayed in
  \eqref{BHKcorrection}, are generated, as said, by states coming from
  D3-D7 and D7-D7 (non-orientable) open strings.\footnote{As we already
    observed in section \ref{ThresCorr}, we do not have winding modes
    in this geometry.} As we have explicitly seen in section
  \ref{ThresCorr}, when combined with the non-perturbative \g
  corrections described above, they lead to \s invariant sets of \g
  corrections for each \a tower. This fact suggests that the sources
  for perturbative \g corrections lift in M-theory to two kinds of BPS
  states \cite{Green:1997di}: 11d supergravitons looping around the
  T-duality circle but carrying zero momentum along the M-theory
  circle (they are in fact bona fide 10d supergravitons) and 11d
  supergravitons possibly carrying units of momentum along the
  T-duality circle but whose worldlines wrap the M-theory circle of
  the F-theory fiber.

  Being still loops of particles around a 1-cycle of the torus fiber,
  these sources should generate corrections proportional to
  \eqref{lMalphaprime} and thus survive the F-theory limit.

\item While 1. and 2. are perturbative in \a, we know that
  non-perturbative \a corrections (for which we do not have an
  explicit expression in section \ref{ThresCorr}) come from Euclidean
  D3-branes wrapped on K3 in type IIB string theory. In F-theory they
  translate to M5 branes over K3$\times T^2$, where $T^2$ is the
  elliptic fiber over a point in the base. They are \s invariant by
  themselves.
\end{enumerate}

\subsubsection{Computation}

Now that we have identified the BPS objects responsible for the \a corrections to the vector multiplet moduli space of F-theory on K3$\times$K3$^\prime$, we can ask ourselves whether we can actually compute these corrections directly in M-theory and then match the result with the full quantum prepotential \eqref{PrepIIB} we already have from type IIB. As anticipated above, a direct Schwinger-loop calculation on K3$^\prime$ of the contributions of 1. and 2., along the lines of \cite{Collinucci:2009nv}, would be hard, due to the non-trivial elliptic fibration. However, we can play with the other F-theory duals of our model to handle this problem (see section \ref{FLifts} for the list of them).

The third F-theory dual is not of great use in our case. It has the advantage though that the K\"ahler potential for vector multiplets of the type IIB theory compactified on  $\tilde{X}_3$ is already exact at the classical level. This is because both the dilaton and the K\"ahler moduli of this theory belong to hypermultiplets. Since the F-theory lift is trivial in this case (no 7-branes), we can immediately extract the holomorphic three-form of the type IIB Calabi-Yau threefold $\tilde{X}_3$ from the holomorphic four-form of the F-theory Calabi-Yau fourfold $\tilde{X}_3\times T^2$: $\Omega_4=\tilde{\Omega}_3\wedge\Omega_1$, with $\Omega_1$ being the unique holomorphic one-form of the torus. Hence we can write our fully corrected K\"ahler potential \eqref{KaehlerPotCorrections}, including the non-perturbative \a corrections, in the compact form 
\be
K(u_k,\bar{u}_k)=-\log i\int_{\tilde{X}_3}\tilde{\Omega}_3\wedge\bar{\tilde{\Omega}}_3\,,
\ee
where $u_k$ are the complex structure moduli of $\tilde{X}_3$, which the K\"ahler moduli of $X_3$ map to via mirror symmetry, and $k=1\ldots,h^{2,1}(\tilde{X}_3)=h^{1,1}(X_3)$. For instance, for the model without Wilson lines, $\tilde{X}_3$ is the mirror of $W\mathbb{P}_{1,1,2,8,12}(24)$ and the moduli $S,T,U$ get mapped to the three complex structure moduli of $\tilde{X}_3$. 

\begin{figure}
  \centering
  \includegraphics[width=\textwidth]{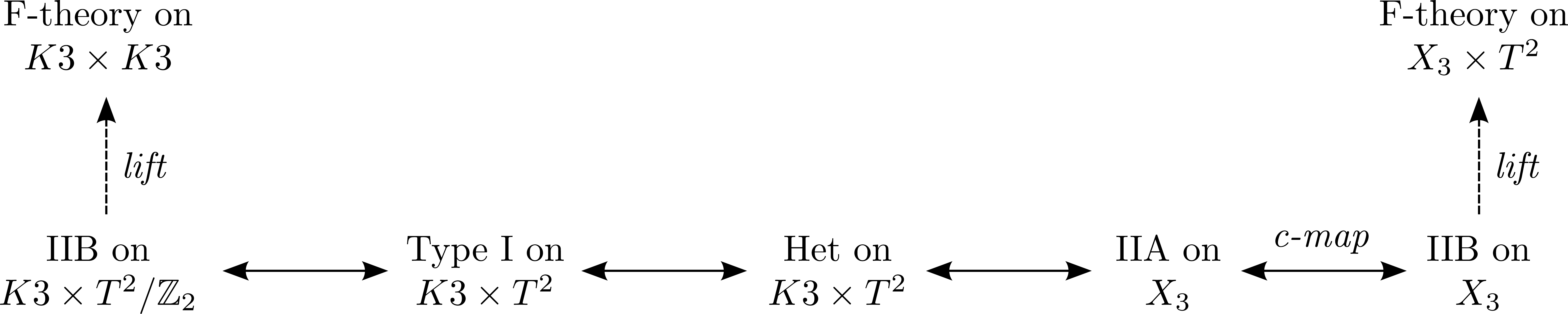}

 \caption{Atlas of dualities used to compute directly in M-theory the quantum corrections to the vector multiplet moduli space of F-theory on K3$\times$K3. In particular, the c-map is used to get a trivial F-theory fibration (top right corner) from a non trivial one (top left corner). The information of the non-trivial fibration is all encoded in the geometry of $X_3$.}
  \label{fig:duality}
\end{figure}

On the other hand a suitable modification of the second F-theory dual allows us to re-compute all the corrections of section \ref{ThresCorr} directly in M-theory, essentially using the method of \cite{Collinucci:2009nv} for a trivial elliptic fibration. Provided the interpretation in section \ref{Sources} of these corrections as \a corrections in F-theory, we thus provide a direct way of computing them, using the M-theory definition of F-theory. The way we get a trivial elliptic fibration out of the non-trivial one characterizing F-theory on\footnote{Recall that $X_3$ admits a non-trivial elliptic fibration and the F-theory torus is understood to be its typical fiber.} $X_3\times T^2$ is based on the so called \emph{c-map} \cite{Ferrara:1989ik}. The c-map basically consists in dimensionally reducing a 4d $\cN=2$ theory on a circle, T-dualizing, and decompactifying the T-dual circle to get another 4d $\cN=2$ theory. Thus it swaps type IIA and type IIB, and the roles of hyper and vector multiplets. 
In our case, we need to apply the c-map to type IIA on $X_3$, which will give us type IIB on $X_3$, which is equivalent to F-theory on the now \emph{trivial} elliptic fibration $X_3\times T^2$. The whole path of dualities we want to follow is summarized schematically in figure \ref{fig:duality}.

Since we are interested in corrections to the vector multiplet moduli
space of the original type IIA, we should be looking at the
hypermultiplet moduli space of the so obtained type IIB. Note that
this operation essentially amounts to swapping the role of the two
tori involved in the second F-theory lift of section \ref{FLifts}: The
elliptic fiber of $X_3$ becomes part of the base and the factorized
$T^2$ becomes the elliptic fiber. In this way, in the F-theory
compactification we end up with in the top right corner of figure
\ref{fig:duality}, the fibration is trivial and thus no 7-branes are
present. We can now safely apply the same method of
\cite{Collinucci:2009nv} to compute corrections. We have to be careful
though, as this procedure is going to give us many more corrections
than we actually need. Since we are dealing with vector multiplets in
type IIA\footnote{For an exhaustive discussion about the
  heterotic/type IIA duality we refer to appendix
  \ref{DualityHet/IIA}.}, we only have \a corrections; those turn via
the c-map into \a corrections to the hypermultiplet metric of the
ensuing type IIB, which \emph{also} admits \g corrections. Therefore,
after the Schwinger-loop computations, we need to extract the tree
level part in \g, and that is going to give us all the \a corrections
of the original F-theory we are aiming for.

Let us describe in detail this computation in the three moduli example, namely the vector multiplet metric of F-theory on K3$\times$K3 in the absence of Wilson lines. As said, the relevant type IIA Calabi-Yau threefold for this model is $X_3=W\mathbb{P}_{1,1,2,8,12}(24)$ (see appendix \ref{DualityHet/IIA} for more details on this duality). Now we are ready to apply the c-map to this type IIA theory and look for quantum corrections to the hypermultiplet metric of the ensuing type IIB on $X_3$. We then compute the latter using the same method as in \cite{Collinucci:2009nv} and we obtain the following prepotential (see also \cite{RoblesLlana:2006is}, where these corrections are computed summing-up $SL(2,\mathbb{Z})$ images)
\bea
\cF_{\rm class}&=&\frac{1}{6}\,\kappa_{\alpha\beta\gamma}\,t_\alpha t_\beta t_\gamma\,,\label{ClassicalF}\\ 
\cF_{\rm pert}&=&-\frac{i}{4(2\pi)^3(\tau_2)^{3/2}}\,\chi(X_3)\,\sum_{(m,n)\neq(0,0)}\frac{\tau_2^{3/2}}{|m\tau+n|^3}\,,\label{PertF}\\ 
\cF_{\rm non-pert}&=&\frac{i}{2(2\pi)^3(\tau_2)^{3/2}}\,\sum_{\bf d}n_{d_\alpha} \sum_{(m,n)\neq(0,0)}\frac{\tau_2^{3/2}}{|m\tau+n|^3}\,e^{2\pi id_\alpha(mc_\alpha+nb_\alpha+i|m\tau+n|j_\alpha)}\,.\label{NonPertF} \nonumber\\ 
\eea
Here, $\kappa_{\alpha\beta\gamma}$ are the classical intersection numbers of $X_3$, $\tau$ denotes the type IIB axiodilaton (complex structure of the factorized torus of the F-theory fourfold), $n_{d_\alpha}$ are the genus-zero Gopakumar-Vafa invariants of $X_3$, and $c_\alpha,b_\alpha,j_\alpha$ are the zero-modes of the RR 2-form, the B-field and the K\"ahler form respectively, expanded along a basis of $H^{1,1}(X_3,\mathbb{Z})$. In \eqref{ClassicalF}, we have the tree-level prepotential, in both \a and \g, of the hypermultiplet sector of type IIB on $X_3$.  Formula \eqref{PertF} gives all the perturbative \a towers of corrections to the prepotential: Each \a tower includes the tree level in \g, a single contribution perturbative in \g and all non-perturbative \g corrections, in a completely $SL(2,\mathbb{Z})$-invariant fashion, much like what we have seen in section \ref{ThresCorr}. Finally, formula \eqref{NonPertF} gives \emph{some} non-perturbative \a corrections, again organized in  $SL(2,\mathbb{Z})$-invariant sets of \g corrections: They are due to euclidean (p,q)-strings wrapping rational curves of $X_3$. It is worth pointing out here that the above expressions do not provide the whole variety of quantum corrections of the hypermultiplet metric of type IIB on $X_3$. The missing corrections are only non-perturbative in \a and are generated by Euclidean D3-branes wrapping divisors of $X_3$ and by the $SL(2,\mathbb{Z})$-invariant set of Euclidean (p,q)-5 branes wrapping $X_3$ itself. 

We are now ready to derive all the corrections to the vector multiplet of the original F-theory on K3$\times$K3. As already stressed, in order to obtain them, we have to select only the tree level part in \g out of the expressions \eqref{ClassicalF},\eqref{PertF},\eqref{NonPertF}. This can clearly be done by sending \g to zero, i.e. $\tau_2\to\infty$. For this purpose, it is useful to recall here the asymptotic behavior of the non-holomorphic Eisenstein series of weight $3/2$ appearing in these quantum corrections
\be\label{EisensteinAsympt}
Z_{3/2}(\tau,\bar\tau)\equiv\sum_{(m,n)\neq(0,0)}\frac{\tau_2^{3/2}}{|m\tau+n|^3}\,\xrightarrow{\tau_2\to\infty}\,2\,\tau_2^{3/2}\zeta(3)\,.
\ee
Hence we obtain:
\begin{itemize}
\item[-] The prepotential in \eqref{ClassicalF} is already tree-level in \g. This gives us directly the polynomial expression in the $S,T,U$ vector moduli of F-theory on K3$\times$K3, as explained in appendix \ref{DualityHet/IIA}.
\item[-] Using \eqref{EisensteinAsympt}, we immediately see that \eqref{PertF} at tree-level in \g becomes
\be
\cF_{\rm pert}\,\xrightarrow{\tau_2\to\infty}\,\frac{i\,\xi}{(2\pi)^3}\,,
\ee
where $\xi$ is defined in \eqref{ConstantTermIIA}. This is exactly the constant term of the corrections to the vector prepotential of F-theory on K3$\times$K3, as can be seen in eq. \eqref{hFunction}.\footnote{Recall that in this case $\chi(X_3)=-480$} As expected, the dependence of \eqref{PertF} on $\tau$ drops out in the tree-level part.
\item[-] Finally, in the limit \g$\to0$, the non-perturbative prepotential \eqref{NonPertF} is going to zero exponentially, unless $m=0$. By using \eqref{EisensteinAsympt} again, we get
\be
\cF_{\rm non-pert}\,\xrightarrow{\tau_2\to\infty}\,\frac{i}{(2\pi)^3}\sum_{\bf d} n_{d_\alpha}{\rm Li}_3(e^{2\pi i t_\alpha d_\alpha})\,,
\ee
where we have defined $t_\alpha\equiv b_\alpha+i\,j_\alpha$, in analogy to \eqref{complexifiedJ}. Again the dependence on $\tau$ drops out at string tree level, as it should. These are the non-perturbative \a corrections of type IIA on $X_3$ and, as explained in appendix \ref{DualityHet/IIA}, they are supposed to include\footnote{Recall from appendix \ref{DualityHet/IIA} that the set of type IIA worldsheet instanton corrections also contains the corrections due to euclidean D3-branes on K3 in the dual F-theory on K3$\times$K3. We do not have an explicit expression for them in section \ref{ThresCorr}.} the infinite series of corrections in \eqref{hFunction} which depend exponentially on the moduli.
\end{itemize}

We have therefore computed the \a corrections to the vector multiplet
moduli space of F-theory on K3$\times$K3 in the absence of 7-brane
moduli $C^i$. If the latter are present, one should consider the type
IIA K3-fibered Calabi-Yau threefold $X_3$ with the right number of
vector moduli ($3$ plus the number of 7-brane moduli switched-on),
regardless of the instanton embedding of the dual $E_8\times E_8$
heterotic string theory, which should not affect the vector multiplet
prepotential.\footnote{ For instance, if we have eight 7-brane moduli
  switched-on in F-theory on K3$\times$K3, we have to consider
  $X_3=W\mathbb{P}_{1,1,12,28,42}(84)$.} Then one computes the quantum
corrections of the hypermultiplet metric of type IIB compactified on
$X_3$ via the Schwinger-loop method developed in
\cite{Collinucci:2009nv}, and finally takes the string tree-level part
of the result sending the type IIB string coupling to zero.

\section{Conclusions}

In this paper we have clarified various aspects of the K\"ahler
potential for vector multiplets in F-theory compactifications on $K3\times K3$. Since we
make heavy use of various known results in the literature, it is
important to clarify which aspects of our analysis are new. As a first
technical result, let us highlight that we explicitly verify that the
weakly coupled limit of \cite{Harvey:1995fq, Henningson:1996jz} agrees with the computation in \cite{Berg:2005ja}, confirming a conjecture in this
last paper. This calculation also explicitly shows the universal
structure of the vector multiplet prepotential in the absence of Wilson lines, as we
obtain the same result for different instanton embeddings.

The calculation in \cite{Harvey:1995fq, Henningson:1996jz} upon which we base our
discussion is done in the context of the heterotic string. By
carefully following the duality map (which we have reviewed in
section~\ref{ReviewModel}), in section~\ref{FTheoryPicture} we
reinterpret these results in the context of F-theory compactified on
K3$\times$K3, and identify the contributing states in the F-theory
language. By analysis of the explicit expressions we also show
explicitly in section~\ref{ThresCorr} that the quantum corrections to
the K\"ahler potential are $SL(2,\mathbb{Z})$ invariant at each
$\alpha'$ level, as one may have expected from the F-theory
picture. (Contributions non-perturbative in $\alpha'$ are missing from
our analysis, it would be interesting to verify explicitly their
behavior under $SL(2,\mathbb{Z})$.) We have also seen that the K\"ahler modulus and the complex structure moduli of the internal manifold indeed mix with each other from the $\alpha'^2$ order, which was not observed at tree-level in $\alpha'$.

We have postponed some of the essential but more technical discussion
of the explicit form of the prepotentials to
appendix~\ref{app:prepotentials}. Appendix~\ref{DualityHet/IIA}
discusses a different dual of the $K3\times T^2$ compactification,
given by type IIA string theory on a Calabi-Yau threefold.

\medskip

We believe that these results are a modest but useful step towards the
ultimate goal of understanding $\alpha'$ corrections to realistic
$\cN=1$ F-theory compactifications. Needless to say, much remains to
be done. Even within the realm of $\cN=2$ compactifications, we have
focused on the easier half of the problem, the vector multiplet moduli
space. Recently there has been some remarkable work on the
understanding of quantum corrections to the hypermultiplet moduli
space
\cite{RoblesLlana:2006is,Alexandrov:2006hx,Saueressig:2007dr,RoblesLlana:2007ae,Alexandrov:2008gh,Pioline:2009qt,Bao:2009fg,Alexandrov:2009zh,Alexandrov:2009qq,Alexandrov:2010np,Alexandrov:2010ca,Alexandrov:2011ac,Alexandrov:2012bu,Alexandrov:2012au,Alexandrov:2012pr}
(see also \cite{Alexandrov:2011va} for a nice review of many of these
results), and it would also be illuminating to carry over these
results to the context of F-theory compactifications, as we have done
here for the vector multiplet moduli space.

More ambitiously, one may wonder how the results in this paper can
shed light on $\cN=1$ compactifications. Since we now have a good
understanding of the physical source of the corrections in the
K3$\times$K3 case, it should be possible to understand at least in
part the structure of the corrections on K3-fibered Calabi-Yau
fourfolds by taking an adiabatic limit. Needless to say, this process
can be rather subtle, but one very important aspect of our analysis is
that it clarifies which kinds of subtleties one finds in going from
$\cN=4$ to $\cN=2$ compactifications (i.e. on going from a trivially
fibered $T^2$ to an elliptically fibered K3). The parent $\cN=4$ theory is given by type IIB compactified on $K3\times T^2$ (or equivalently F-theory on $K3\times T^4$) with no 7-branes. The classical K\"ahler potential is given by the sum of $\cK_K$ and $\cK_c$ in \eqref{KkKc2}, where now   $K3^\prime\to T^2\times T^2$ and thus $\int\Omega_2^\prime\wedge\bar\Omega_2^\prime\to(S-\bar S)(U-\bar U)$. Quantum mechanically, since $\chi(K3\times T^2)=0$, there are neither perturbative \g corrections nor those non-perturbative due to D(-1)-instantons (i.e. \eqref{PertF} vanishes identically). Moreover, on the one hand non-perturbative \a corrections due to euclidean D3 branes are generically all there (both the ones which correct the vector multiplets and the ones which correct the hypermultiplets of our $\cN=2$ theory). On the other hand,  F1/D1-instantons and NS5/D5-instantons, which are absent in our $\cN=2$ theory due to the orientifold projection, do induce non-perturbative \a corrections to the K\"ahler potential of the parent $\cN=4$ theory. 

Hopefully this pattern can serve as a guide in going from the $\cN=2$ theory arising from F-theory on $CY_3\times T^2$ to the $\cN=1$ theory arising from F-theory on non-trivially fibered $CY_4$. 
One possible route would be trying to reproduce the corrections we discussed in this paper by making the computation of the graviton loop processes directly on the non trivial elliptic fibration. Succeeding in this aim would mean having at hand a concrete technique to apply to the more complicated non-trivial fibrations involved in $\cN=1$ compactifications of F-theory.

Another possible avenue of research would be to study vacua that
spontaneously break the $\cN=2$ symmetry down to $\cN=1$
\cite{Ferrara:1995gu,Ferrara:1995xi,Fre:1996js,Louis:2009xd,Louis:2010ui}. In
this class of scenarios one can sometimes obtain interesting
information about the $\cN=1$ dynamics starting from the $\cN=2$
theory \cite{Taylor:1999ii,Lawrence:2004zk}. For instance, the spontaneous breaking can be generated by fluxes, which in turn, in some cases, induce warping. The effects of the latter on the K\"ahler potential are analyzed in \cite{Marchesano:2008rg, Marchesano:2010bs}, where the authors also provide the lift of the type IIB results to F-theory.

Finally, it would be interesting to see if it is possible to reproduce
the effects that we have found from a higher derivative modification
of the 11d action, similarly to the effect in the trivial fibration
case analyzed in \cite{Green:1997tv,Green:1997di,Green:1997as}
(although possibly with a different number of graviton insertions).

\medskip

These are all interesting questions, and we hope to return to them in
the near future.

\acknowledgments

We would like to thank Andr\'es Collinucci, Ioannis Florakis, Thomas Grimm, Michael
Haack, Kenji Hashimoto, Stefan Hohenegger, Ling-Yan Hung, Pablo
Soler and Roberto Valandro for illuminating discussions. HH, I.G.-E. and RS would like to
thank the Hong Kong Institute for Advanced Study at HKUST, where this
project was born, for hospitality and financial support, and the Bethe
Center for Theoretical Physics at the University of Bonn for
hospitality during the last stages of this work. HH would also like to thank Max-Planck Institute for Physics for hospitality and financial support during a part of this work. GS was supported in
part by DOE grant DE-FG-02-95ER40896. GS would also like to thank the
University of Amsterdam for hospitality during part of this work, as
he was visiting the Institute for Theoretical Physics as the Johannes
Diderik van der Waals Chair. I.G.-E. would also like to thank
N.~Hasegawa for kind encouragement and support.

\appendix

\section{Explicit expressions of the prepotential}
\label{app:prepotentials}

\subsection{Generalities}

In this paper we have discussed the vector multiplet prepotential of type \Ip string theory which was obtained by the chain of dualities starting from the prepotential of $SO(32)$ heterotic string theory on $K3 \times T^2$. The general form of the prepotential has been calculated in \cite{Henningson:1996jz}. Ref.~\cite{Henningson:1996jz} obtained the explicit form of the prepotential of the $E_8 \times E_8$ or $SO(32)$ heterotic string theory on $T^4/\bbZ_n \times T^2$ with $n= 2, 3, 4, 6$ including general Wilson lines. Hence, one can apply the result in \cite{Henningson:1996jz} to compute the prepotential of any theory obtained from the heterotic string theory on $T^4/\bbZ_n \times T^2$. In particular, we have studied the prepotential of type \Ip string theory which is dual to a particular type of the BSGP model. That BSGP model has a dual heterotic description realized by an $SO(32)$ heterotic string theory on $T^4/\bbZ_2 \times T^2$.
Since our analysis used the explicit form of the prepotential, we show in this appendix the computation of the prepotential of the theory by utilizing the result in \cite{Henningson:1996jz}. 

Before going to the specific example, let us summarize the general result of \cite{Henningson:1996jz}. We let the orbifold group $Z_n$ act on the complex coordinates $z^1, z^2$ of $T^4$ as 
\be
z^1 \rightarrow e^{2\pi i a/n}z^1,\qquad z^2 \rightarrow e^{-2 \pi i a /n}z^2,
\ee
where $a = 0, \cdots, n-1$. In the bosonic formulation of heterotic string theory, the gauge degrees of freedom are described by sixteen left-moving bosons and they take their values in the $E_8 \times E_8$ root lattice or the $Spin(32)/Z_2$ weight lattice. From the $T^2$ compactifications we also have two left-moving bosons and two right-moving bosons. In total, those fields take their values in an even self-dual lattice $\Gamma^{18,2}$, which can be taken to be $\Gamma^{16,0} \oplus \Gamma^{1,1} \oplus \Gamma^{1,1}$. Here $\Gamma^{16,0}$ represents either $E_8 \times E_8$ root lattice or $Spin(32)/Z_2$ weight lattice. To cancel the space-time anomaly, we embed the orbifold action in the gauge degrees of freedom by a shift $\frac{a}{n}\gamma$ where $\gamma \in \Gamma^{18,2}$. 

The moduli of the theory are described by the $O(18,2, \mathbb{R})$
rotations which produce inequivalent even self-dual lattices
$\Gamma^{18,2}$. After removing the redundancy which gives equivalent
even self-dual lattices, the classical moduli space is the quotient by
the discrete T-duality group of
\begin{align}
  O(2,18, \mathbb{R})/(O(2, \mathbb{R}) \times O(18, \mathbb{R})).
\end{align}
The space could be parameterized by complex moduli\footnote{The bar on
  a letter always indicates that the element represented by the letter
  is in $\Gamma^{16,0}$.} $(\bar{y}, y^{+}, y^{-})$ where $y^{-}_2 >
0$ and $(y_2, y_2) < 0$ \cite{Ceresole:1994cx, deWit:1995zg}. Here the subscript $2$ represents an imaginary part. The inner product is
defined as
\begin{align}
(y, y^{\prime}) = \bar{y}\cdot \bar{y}^{\prime} -
y^{+}y^{-\prime} - y^{-}y^{+\prime}.
\end{align} The classical K\"ahler potential for the moduli $y$ is
written by means of the inner product
\begin{align}
  K_{\text{classical}} = -
  \log \left( -(y_2, y_2) \right).\label{classical.kahler}
\end{align}

For later convenience we introduce some further notations. We define $\bar{R} = \bar{r} + \frac{a}{n}\bar{\gamma}$ where $\bar{r}, \bar{\gamma} \in \Gamma^{16,0}$. Also, we introduce $R = (\bar{R}, -l, -k)$ and define the positivity of $R$ as 
\be
k > 0,\;\;{\rm or} \;\; k=0, \; l > 0,\;\;{\rm or}\;\; k=l=0, \; \bar{R} > 0\,,
\ee
where $\bar{R}$ belongs to the weight lattices of various representations of the gauge group left unbroken by the orbifold shift. Its positivity is defined by dividing each lattice in sets of positive and negative weights, and usually this is conventionally done by declaring positive a weight vector whose first non-zero component is positive (see appendix \ref{app:SpecificWL} for an explicit example).
Furthermore, we denote by $Q$ some generator of a simple factor in the gauge group. Modular invariance of the torus partition function requires
\be
\bar{y}\cdot \bar{Q} = 0. \label{req.Q}
\ee

By using the notation above, one can express the exact prepotential of heterotic string theory on $T^4/\bbZ_n \times T^2$. We focus on a fundamental chamber of the moduli space 
\bea
&&0 < \frac{\bar{R}\cdot \bar{y}_2}{y^{-}_2} < 1 \quad {\rm for}\;\;\bar{R} > 0, \;\bar{R}\cdot\bar{R} \leq 2,\label{weyl1}\\
&&0 <  y^{-}_2 < y_2^{+}.\label{weyl2}
\eea
When there are no Wilson lines, the fundamental Weyl chamber is characterized only by \eqref{weyl2}. The fundamental Weyl chamber \eqref{weyl1} and \eqref{weyl2} may be understood by the convergence of the series appearing in the prepotential. The final expression of the prepotential is then
\be
h(y) = -\frac{1}{32\pi}(d_{{\rm gauge}})_{ABC}y^A y^B y^C - i\frac{\zeta(3)\chi}{8\pi^4} - \frac{i}{16 \pi^4} \sum_{R > 0}{}^{\prime} d(R) {\rm li}_3((R,y)),
\label{general}
\ee
where the prime on the sum of the third term means that $k=l=0$ and $\bar{R}\cdot\bar{y}=0$ for generic values of the moduli are omitted, and 
\be
{\rm li}_3(x) = {\rm Li}_3(e^{2\pi i x}) = \sum_{p=1}^{\infty}\frac{(e^{2\pi i x})^p}{p^3}.
\ee

The coefficients of \eqref{general} are also explicitly computed, and $(d_{{\rm gauge}})_{ABC}$ is
\bea
(d_{{\rm gauge}})_{ABC}y_2^A y_2^B y_2^C &=& \sum_{\bar{r},a}d(\bar{R})\Big(\frac{(\bar{R}\cdot \bar{Q})^2}{\bar{Q} \cdot \bar{Q}}(-2 |\bar{R} \cdot \bar{y}_2| \bar{y}_2 \cdot \bar{y}_2 + 4|\bar{R}\cdot \bar{y}_2|y_2^+ y_2^- + \frac{2}{3}\bar{y}_2\cdot\bar{y}_2y^+_2 \nonumber\\
&& -4(\bar{R}\cdot\bar{y}_2)^2y_2^+ + \frac{1}{3}\bar{y}_2\cdot\bar{y}_2y_2^- -\frac{2}{3}y_2^+y_2^-y_2^- - \frac{2}{3}y_2^+y_2^+y_2^-) \nonumber\\
&&+\frac{2}{3}|\bar{R} \cdot \bar{y}_2|^3 - \frac{1}{30}\bar{y}_2 \cdot \bar{y}_2 y_2^+ - \frac{1}{3}(\bar{R}\cdot\bar{y}_2)^2y_2^- + \frac{1}{30}y_2^+y_2^+y_2^-+\frac{1}{90}y_2^-y_2^-y_2^-\Big).\nonumber\\ \label{gauge}
\eea
$d(R)$ in \eqref{general} is defined as 
\be
d(R) = \sum_b \frac{1}{n}e^{2\pi i \frac{b}{n} \bar{R} \cdot \bar{\gamma}}c_{a,b}\left(-\frac{1}{2}(R, R)\right),
\label{dR}
\ee
where $c_{a,b}(h)$ is an expansion coefficient of 
\be\label{ExpansCoefficients}
e^{-2\pi i \frac{ab}{n^2} \gamma^2} \eta^{-20}(\tau) Z_{a,b}^{K3}(\tau) = \sum_{h \geq -1}c_{a,b}(h)q^h,
\ee
and $b=0,1, \cdots, n-1$. Here $Z^{K3}_{a,b}(\tau)$ is 
\be
Z^{K3}_{a,b}(\tau) = k_{a,b}q^{-\left(\frac{a}{n}\right)^2}\eta^2(\tau)\Theta_1^{-2}\left(\tau\frac{a}{n}+\frac{b}{n} | \tau \right),
\label{partitionk3}
\ee
where $\eta(\tau)$ and $\Theta_1(\nu | \tau)$ are the Dedekind eta function and the Jacobi theta function respectively. The constant $k_{a,b}$ in \eqref{partitionk3} is 
\be
k_{0,b} = 64 \sin^4 \pi\frac{b}{n}
\ee
for $a=0$ and 
\be
\frac{k_{a,b}}{k_{a,a+b}} = e^{i \pi \frac{a^2}{n^2}(2-\gamma^2)},\qquad \frac{k_{a,b}}{k_{b,-a}} = e^{-2\pi i \frac{ab}{n^2}(2-\gamma^2)}
\ee
for $a \neq 0$. Finally $\chi$ in \eqref{general} is 
\be
\chi = \frac{1}{4}\sum_{\bar{r}, a}{}^0 d(\bar{R}),
\label{chi}
\ee
where the superscript $0$ indicates that the sum is only for $a$ such that $\bar{R}\cdot\bar{y}=0$ for generic values of the moduli $\bar{y}$ for a given $\bar{r}$.  

The cubic terms \eqref{gauge} seems to depend on the choice of the generator $Q$ of a simple gauge group factor. However, \eqref{gauge} is independent of $Q$ up to a term $\sum_i(y_2, y_2)c^i y_i$ for some real constant $c^i$, by taking into account the constraint on $Q$ \eqref{req.Q}. The term $\sum_i(y_2, y_2)c^i y_i$ can be reabsorbed in a shift of $S$ in the classical prepotential and hence it does not give any physical effect in the low energy effective theory.

\subsection{A universal prepotential without Wilson lines}\label{app:UniversalPrepotential}

We now turn to a specific example, namely the heterotic string theory
on $T^4/\bbZ_2 \times T^2$ with a maximal $SU(16) \times U(1)^4$ gauge
group\footnote{The $SU(16)$ gauge symmetry can be completely higgsed
  by the vev of the charged hypermultiplets.} which is dual to a
particular BSGP model. We have three scalars $S_H, T_H, U_H$ in the
three vector multiplets associated to three of the four $U(1)$
symmetries, the other $U(1)$ being associated to the photon in the
$\mathcal{N}=2$ supergravity multiplet. In the type I language, the
SO(32) gauge group present on the 32 D9-branes gets broken by the
orbifold shift down to $U(16)$. The center-of-mass $U(1)$ symmetry, in
turn, is broken by a non-perturbative effect
\cite{Berkooz:1996iz}. Moreover, at each fixed point in $T^4/\bbZ_2$,
there is a ``half''-5-brane. These are the type I duals of the
heterotic small instantons \cite{Witten:1995gx}. Each carries charge
1/2 and, in the model we are focusing on, they exhaust, together with
16 units of non-vector instantons on the singularities, the total
instanton number of 24. The sixteen half-5-branes produce $U(1)^{16}$
gauge bosons which get massive due to St\"uckelberg couplings
\cite{Berkooz:1996iz,Aldazabal:1997wi}. By T-dualizing to type \Ip, we
thus find 16 space-filling half-D3-branes, which have the total charge
of 8 and moreover are completely stuck in all the internal directions
(i.e. they have no deformation moduli at all). To cancel the remaining
16 units of the gravitational D3-tadpole we have an appropriate flux
background on the D7-branes, which is induced by the instanton without
vector structure.\footnote{In the generic case one has just a smooth
  instanton bundle with instanton number 24 on a smooth K3. In this
  case everything is higgsed, no D3-branes are present and the
  gravitational tadpole is canceled only by fluxes.}

Let us first focus on the case without Wilson lines. The model can be obtained in heterotic string theory by considering a special embedding of the orbifold action, namely we take \cite{Aldazabal:1997wi} 
\be
\bar{\gamma} = \frac{1}{2}(1, \cdots, 1, -3) \in \Gamma^{16,0}.
\label{gamma}
\ee
The orbifold embedding breaks the gauge group $SO(32)$ into $SU(16)
\times U(1)$. Since we also turn off all the Wilson lines, we set
$\bar{y}=0$. Therefore, the prepotential can be written only in terms of two
moduli, $y^{+}, y^{-}$. Those two moduli correspond to the
complexified K\"ahler modulus $T_H$ and the complex structure modulus $U_H$
of the torus $T^2$. We choose them as $(T_H, U_H) = (y^{+}, y^{-})$. Then, the
classical K\"ahler potential \eqref{classical.kahler} is 
\be
K_{\text{classical}} = -\log\left( 2(T_H)_2(U_H)_2 \right).
\ee

There is also an axio-dilaton modulus $S_H$ which is the scalar component of another vector multiplet. The full classical K\"ahler potential is 
\be
K_{\text{classical}} = -\log\left(\alpha (S_H)_2\right) -\log\left( 2(T_H)_2(U_H)_2 \right), \label{kahler1}
\ee
where $\alpha$ is a real constant and depends on the normalization of $S$. The prepotential which reproduces \eqref{kahler1} is 
\be
\mathcal{F}_{\text{classical}} = -\frac{\alpha}{4} S_HT_HU_H. \label{cl.pre}
\ee

Let us first compute the coefficients $(d_{{\rm gauge}})_{ABC}$ of the cubic terms in \eqref{general}. After setting $\bar{y}=0$, $(d_{{\rm gauge}})_{ABC}y^A y^B y^C$ becomes
\be
(d_{{\rm gauge}})_{ABC}y^A y^B y^C = \sum_{\bar{r}, a}\frac{d(\bar{R})}{90}U_H^3 + \beta\, T_HU_H^2 + \gamma\, T_H^2 U_H
\ee
with some constants $\beta, \gamma$. 
In fact, the terms like $\beta T_HU_H^2 + \gamma T_H^2U_H$ can be absorbed by the shift of $-\frac{\alpha}{4} S_H \rightarrow -\frac{\alpha}{4} S_H + \beta U_H + \gamma T_H$ in the classical prepotential \eqref{cl.pre}. Hence, the coefficients of the cubic term without the redundancy is 
\be
(d_{{\rm gauge}})_{ABC}y^A y^B y^C = \sum_{\bar{r}, a}\frac{d(\bar{R})}{90}U_H^3. 
\label{gauge1}
\ee

We move on to the computation of $d(\bar{R})$. Although the calculation of $d(\bar{R})$ in \eqref{gauge} needs the summation over the $\bar{r} \in \Gamma^{16,0}$, it turns out that only a finite number of $\bar{r}$ contributes to \eqref{gauge}. We first compute the contributions from the untwisted modes, namely $a=0$. When $a=0$, $(\bar{R}, \bar{R}) = \bar{r}\cdot\bar{r}$ and the non-zero contribution should come from $\bar{r}$ with $\bar{r}\cdot\bar{r} \leq 2$. Therefore, those $\bar{r}$s are $\bar{r} = 0$ or roots of $SO(32)$. Hence the $a=0$ part of \eqref{gauge1} is 
\be
\sum_{\bar{r}}d(\bar{R}) = \frac{1}{2}c_{0,1}(0) + \frac{1}{2}\sum_{\bar{r} \in {\rm roots\;of\;SO(32)}} e^{\pi i \bar{r}\cdot \gamma} c_{0,1}(-1).
\label{gauge1-1}
\ee 
Here, we used fact that $c_{0,0}(h) = 0$ since the constants $k_{a,b}$ with $n=2$ and the $\gamma$ of \eqref{gamma} are
\be
k_{0,0} = 0, \quad k_{0,1} = 64 , \quad k_{1,0}= 64, \quad k_{1,1}= -64.
\ee
Since $c_{0,1}(h)$ is the expansion coefficient of 
\be
\sum_{h \geq -1} c_{0,1}(h)q^h = 64 \eta^{-18}(\tau) \Theta_{1}^{-2}\left(\frac{1}{2} | \tau \right), 
\ee
 we have $c_{0,1}(-1) = 16,\; c_{0,1}(0) = 256$. Inserting these results into \eqref{gauge1-1}, we get
\be
\sum_{\bar{r}}d(\bar{R}) = 256/2 + 240\times 16/2 - 240 \times 16/2 = 128.\label{a0}
\ee
Note that the contributions from the roots of $SO(32)$ are canceled with each other in \eqref{a0}. 
 
We also have a contribution from $a=1$ part to \eqref{gauge1}, which corresponds to the contributions from the twisted modes. Eq.~\eqref{dR} for $a=1$ becomes\be
d(\bar{R}) = \frac{1}{2}c_{1,0}\left(-\frac{1}{2}\Big(\bar{r} + \frac{1}{2}\gamma\Big)^2\right) - \frac{1}{2}e^{\pi i \bar{r} \cdot \gamma}c_{1,1}\left(-\frac{1}{2}\Big(\bar{r} + \frac{1}{2}\gamma\Big)^2\right). \label{dR2}
\ee
Here $c_{1,0}(h)$ and $c_{1,1}(h)$ are expansion coefficients of 
\bea
\sum_{h \geq -1}c_{1,0}(h)q^h &=& 64 \eta^{-18}(\tau) q^{-\frac{1}{4}}\Theta^{-2}_1\left(\frac{1}{2}\tau | \tau\right) = -64\eta^{-18}(\tau) \Theta_{4}^{-2}(0 | \tau),\label{theta2}\\
\sum_{h \geq -1}c_{1,1}(h)q^h &=& 64 \eta^{-18} q^{-\frac{1}{4}} \eta^{-18}(\tau) \Theta_{1}^{-2}\left(\frac{1}{2}\tau + \frac{1}{2} | \tau\right) = 64 \eta^{-18}(\tau)\Theta_{3}^{-2}( 0 | \tau).\label{theta3}
\eea
Both series \eqref{theta2} and \eqref{theta3} start from $q^{-3/4}$ and the power of each term is $-\frac{3}{4} + \frac{1}{2}\mathbb{Z}_{\geq 0}$. Hence, the possibility for non-zero coefficients is $-\frac{1}{2}\Big(\bar{r} + \frac{1}{2}\gamma\Big)^2 = -\frac{3}{4}$ or $-\frac{1}{2}\Big(\bar{r} + \frac{1}{2}\gamma\Big)^2 = -\frac{1}{4}$. The elements in $\Gamma^{16,0}$ which satisfy the above
former equation are 
\bea
\bar{r} &=& 0,\label{first}\\
\bar{r} &=& -e_i + e_{16},\quad (i=1, \cdots, 15),\label{second}\\
\bar{r} &=& \pm \frac{1}{2}e_1 + \cdots + \pm\frac{1}{2}e_{15} + \frac{1}{2}e_{16},\label{3rd}\\
\bar{r} &=& -\gamma \label{fourth},
\eea
where only one sign of \eqref{3rd} has to be plus and all the other minus. $e_1, \cdots, e_{16}$ are orthonormal bases of the sixteen-dimensional space $\mathbb{R}^{16}$. Note that the sum $\bar{r} + \frac{1}{2}\bar{\gamma}$ for all the weights \eqref{first}--\eqref{fourth} are expressed as 
\be
\pm \left(\frac{1}{4}, \cdots, \frac{1}{4}, -\frac{3}{4}, \frac{1}{4}, \cdots, \frac{1}{4} \right). \label{fifth}
\ee
where only one component in the sixteen dimensional vector is $-\frac{3}{4}$. The weights \eqref{fifth} can be also expressed as 
\be
\pm e_i \mp\left(\frac{1}{4}, \cdots, \frac{1}{4} \right), (i=1,\cdots, 16). \label{sixth}
\ee
Therefore, the weight \eqref{sixth} may be understood as the fundamental or the anti-fundamental weight of the $SU(16)$ since $\left(\frac{1}{4}, \cdots, \frac{1}{4}\right)$ is a singlet under the $SU(16)$. On the other hand, there are no elements in $\Gamma^{16,0}$ which satisfy $-\frac{1}{2}\Big(\bar{r} + \frac{1}{2}\gamma\Big)^2 = -\frac{1}{4}$. Then, the sum of \eqref{dR2} over the elements in $\Gamma^{16,0}$ is 
\bea
\sum_{\bar{r}} d(\bar{R}) &=& \frac{1}{2}c_{1,0}\left(-\frac{3}{4}\right) \times 32 - \frac{1}{2}c_{1,1}\left(-\frac{3}{4}\right) \times 32\\
&=& - 64 \times 32,\label{a1}
\eea 
where we used $c_{1,0}\left(-\frac{3}{4} \right) = -64, c_{1,1}\left(-\frac{3}{4}\right) = 64$.

Summarizing \eqref{a0} and \eqref{a1}, we finally obtain
\be
\sum_{\bar{r}, a}d(\bar{R}) = 128 - 64 \times 32 = -1920.\label{sum1}
\ee
Therefore, the net non-zero contributions to \eqref{sum1} come from the twisted modes at fixed points. Finally, \eqref{gauge1} is 
\be
(d_{{\rm gauge}})_{ABC}y^A y^B y^C = -\frac{64}{3} U_H^3.
\ee

The computation of \eqref{chi} is also performed in a similar way. Since we turn off the Wilson lines $\bar{y} = 0$, one always has $\bar{R} \cdot \bar{y} = 0$. Therefore, the summation in \eqref{chi} is exactly the same as the summation in \eqref{gauge1}. The final result is 
\be
\chi = \frac{-1920}{4} = -480.
\ee

The remaining term is the third term in \eqref{general}. Note that the prime in the sum of the third term in \eqref{general} in this case indicates that there is no contribution from $k=l=0, \bar{R} > 0$. Then, let us compute the coefficients of a few terms with $kl \leq 0$ as examples. Since $-\frac{1}{2}\bar{R}\cdot\bar{R} + kl \geq -1$ due to eq. \eqref{ExpansCoefficients}, the lowest value for $kl$ is $-1$. In the fundamental chamber, there is only one term with $kl=-1$, namely $k=1, l=-1$. In this case, $\bar{R} = 0$ and we have
\be
\sum_{R > 0}{}^{\prime}d(R) \rightarrow \frac{1}{2} c_{0,1}(-1) = 8 \quad {\rm for\;the\;term\;with\;}kl=-1.
\ee
Next, we consider the terms with $kl=0$. Then, the constraint for $\bar{R}$ is $-\frac{1}{2}\bar{R}\cdot\bar{R}\geq -1$. Since $-\frac{1}{2}\bar{R}\cdot\bar{R} \leq 0$, the non-zero contributions come from $\bar{r}\cdot\bar{r} = 2$ or $\bar{r} = 0$ for $a = 0$ and $-\frac{1}{2}\left(\bar{r} + \frac{1}{2}\gamma\right)^2 = -\frac{3}{4}, -\frac{1}{4}$ for $a=1$. Hence, the total contribution is the same as \eqref{sum1} and we have
\be
\sum_{R > 0}{}^{\prime} d(R) \rightarrow -1920 \quad {\rm for\;the\;terms\;with\;}kl=0
\ee

Summarizing all the results we computed, we finally obtain the explicit expression for the quantum prepotential of the heterotic string theory which is dual to the BSGP model without Wilson lines,
\bea
h(T_H, U_H)&=& \frac{2}{3\pi}U_H^3 + i\frac{60 \zeta(3)}{\pi^4} \nonumber \\
&& - \frac{i}{(2\pi)^4}\Big( 8{\rm Li}_3(e^{2\pi i(T_H-U_H)}) - 1920 \sum_{l > 0}{\rm Li}_3(e^{2\pi ilU_H}) - 1920 \sum_{k > 0}{\rm Li}_3(e^{2\pi ikT_H}) + \cdots\Big).\nonumber \\ \label{explicit1}
\eea
In the chamber $(T_H)_2 < (U_H)_2$, then the prepotential is the same one as \eqref{explicit1} with $T_H$ and $U_H$ exchanged. The full prepotential is the sum of \eqref{cl.pre} and \eqref{explicit1}.

One can compare the result \eqref{explicit1} with the prepotential of $E_8 \times E_8$ heterotic string theory on $T^4/\bbZ_2 \times T^2$ in the standard embedding. Certainly, the prepotential can be obtained from the general form \eqref{general}, but Ref.~\cite{Harvey:1995fq} has a nicer expression for it. In the absence of Wilson lines, the explicit form of the quantum prepotential is 
\be
h^{HM}(T_H, U_H) = \frac{U_H^3}{12 \pi} + \frac{15 i}{2\pi^4}\zeta(3) - \frac{i}{(2\pi)^4}\left[ {\rm Li}_3(e^{2\pi i(T_H-U_H)}) + \sum_{\begin{subarray}{c} k,l\geq 0 \\ (k,l)\neq 0 \end{subarray}} c(kl) \,{\rm Li}_3(e^{2\pi i(kT_H + lU_H)})\right], \label{HM}
\ee
where
\be
\sum_{n=-1}^{\infty}c(n)q^n = \frac{E_6 E_4}{\eta^{24}}(q) = \frac{1}{q} - 240 + \cdots. \label{eisenstein}
\ee
$E_{6,4}$ are the Eisenstein series. By multiplying \eqref{HM} by eight, 
then one can see the exact matching with \eqref{explicit1},
\be
h(T_H,U_H) = 8h^{HM}(T_H, U_H), \label{normalization1}
\ee
at least with respect to the terms explicitly written in \eqref{explicit1}. 

For the sake of completeness, the classical K\"ahler potential in \cite{Harvey:1995fq} is 
\be
K_{\text{classical}}^{HM} = -\log\left(8(S_H^{HM})_2\right) - \log\left(-(y_2, y_2)\right). \label{kahlerHM}
\ee
Therefore, the classical prepotential which reproduces \eqref{kahlerHM} is 
\be
\mathcal{F}_{\text{classical}}^{HM} = -2S_H^{HM}T_HU_H. \label{cl.pre2}
\ee
By taking into account the matching \eqref{normalization1}, one may identify the normalization of $S^{HM}_H$ with 
\be
S^{HM}_H = \alpha S_H.
\ee

In fact, it is expected that the prepotential \eqref{explicit1} exactly matches with \eqref{HM}. In order to see that, let us first see the duality between the $SO(32)$ heterotic string theories and the $E_8 \times E_8$ heterotic string theories. The $SO(32)$ heterotic string model which is dual to the BSGP model is conjectured to be dual to the $E_8 \times E_8$ heterotic string theory on $K3 \times T^2$ with the symmetric instanton embedding $(12, 12)$ \cite{Berkooz:1996iz}. At a generic point in the hypermultiplet moduli space, the vev of the charged hypermultiplet moduli can completely break the non-Abelian gauge symmetry on both sides. On those points, we only have $U(1)^{3+1}$ gauge symmetries for both theories and they have the same prepotential for the three vector multiplet moduli. Note that the explicit form of the prepotential would also be the same as \eqref{explicit1} since one can move to a smooth K3 surface by varying the neutral hypermultiplet moduli which does not affect the vector multiplet moduli space.    

Next, we will argue that the prepotential of the $E_8 \times E_8$ heterotic model in the standard instanton embedding (24,0) without Wilson lines is the same as that of the $E_8 \times E_8$ heterotic string model with the instanton embedding $(12, 12)$. Recall that the prepotential \eqref{general} was obtained by solving the differential equation \cite{Henningson:1996jz}
\bea
&&{\rm Re}\left[-\frac{1}{s+4}\frac{\partial}{\partial y}\cdot \frac{\partial}{\partial y}(-ih(y)) + \frac{1}{(y_2, y_2)}\left( -ih(y) - iy_2^A \frac{\partial}{\partial y^A}(-ih(y))\right) \right]\nonumber \\
&=&\frac{1}{16\pi^2}\Delta_{{\rm gauge}} + \frac{1}{2(s+4)\pi^2}{\rm Re}\;{\rm log}\Psi_{{\rm gauge}} + \frac{b_{{\rm gauge}}}{16\pi^2}(-(y_2, y_2)) + {\rm const}, \label{diff}
\eea
where $s$ is the number of Wilson lines, $\Delta_{{\rm gauge}}$ is the gauge threshold correction in string theory and $b_{{\rm gauge}}$ is the coefficient of the $\beta$ function for the gauge coupling. Our case corresponds to $s=0$.

The differential equation \eqref{diff} was obtained by comparing the string theoretic gauge coupling with the field theoretic gauge coupling. The last constant term in \eqref{diff} is due to the fact that the classical K\"ahler potential appears in the expression of the field theoretic gauge coupling.  The K\"ahler potential is defined up to a sum of a holomorphic function and an anti-holomorphic function, and hence we can choose the constant term as we like. The gauge threshold correction $\Delta_{\text{gauge}}$ has an ambiguous constant term which is subject to an infrared regularization. We choose the constant term in \eqref{diff} such that it precisely cancels the constant term coming from  $\Delta_{\text{gauge}}$.

In fact, the general formula for $\Psi_{{\rm gauge}}$ in the case without Wilson lines is \cite{deWit:1995zg}
\be
 \frac{1}{8\pi^2}{\rm log}\Psi_{{\rm gauge}} = \frac{b_{{\rm gauge}}}{4\pi^2}{\rm log}[\eta(T_H)\eta(U_H)] + \frac{1}{8\pi^2}{\rm log}(j(T_H) - j(U_H)).\label{gaugemodular}
\ee
Furthermore, the gauge threshold correction $\Delta_{{\rm gauge}}$ without Wilson lines has also a universal structure \cite{Kiritsis:1996dn} and can be written as
\be
\Delta_{{\rm gauge}} = b_{{\rm gauge}}\Delta + ({\rm universal}), \label{universal}
\ee
where the $({\rm universal})$ part only depends on the difference between the number of hypermultiplets ($n_H$) and the vector multiplets ($n_V$). As for the $E_8 \times E_8$ heterotic string without Wilson lines in the standard embedding and the $E_8 \times E_8$ heterotic string with the $(12,12)$ embedding, both have $n_H - n_V = 240$. The $\Delta$ in \eqref{universal} is 
\bea
\Delta &=& \int_{\mathcal{F}}\frac{d^2\tau}{\tau_2}[\Gamma^{2,2}(T_H, U_H) - 1]\nonumber\\
&=&-{\rm log}(|\eta(T_H)|^4|\eta(U_H)|^4(T_H)_2(U_H)_2) + {\rm const}.\label{torusthreshold}
\eea
The constant term in \eqref{torusthreshold} represents the ambiguity of the infrared regularization. Then, by inserting \eqref{gaugemodular} and \eqref{universal} into the right hand side of \eqref{diff}, one gets
\be
\frac{1}{16\pi^2}({\rm universal}) + \frac{1}{8\pi^2}{\rm Re}[{\rm log}j(T_H) - j(U_H)] + {\rm const}.
\ee

Therefore the differential equations for the prepotentials are completely the same for the $E_8 \times E_8$ heterotic string without Wilson lines in the standard embedding and the $E_8 \times E_8$ heterotic string with the $(12,12)$ embedding. Hence, one may conclude that the prepotential of the $E_8 \times E_8$ heterotic string without Wilson lines in the standard embedding is the same as that of the $E_8 \times E_8$ heterotic string with the instanton embedding $(12,12)$. Based on this argument, we will use the expression \eqref{HM} for the prepotential of the $SO(32)$ heterotic string model which is dual to the BSGP model. 

\subsection{Inclusion of specific Wilson lines}\label{app:SpecificWL}

The explicit prepotential \eqref{general} is general enough to consider the case with Wilson line moduli. For the description of the inclusion of the Wilson line moduli, we focus on a region in the hypermultiplet moduli space where we have the whole $SU(16)$ gauge group. The $SU(16)$ gauge group will be broken by turning on Wilson line moduli. Aiming for the comparison with the result in \cite{Berg:2005ja}, we turn on a Wilson line in the $SU(16)$ gauge group such that the Wilson line moduli have the following form
\be
\bar{y} = (A^1_H, \cdots, A^1_H, \cdots, A^k_H, \cdots, A^k_H).
\ee 
where the number of $A^i_H$ is $N_i$ and $\sum_{i=1}^{k}N_i = 16$. We also have a relation 
\be
\sum_{i=1}^{k} A^i_H N_i = 0 \label{speciality}
\ee
due to the tracelessness condition on the generators of $SU(16)$.

Let us first consider the classical prepotential. The classical K\"ahler potential \eqref{classical.kahler} including the axio-dilaton moduli is 
\be
K_{\text{classical}}= -\log\left(\alpha (S_H)_2 \right) - \log\left(2(\hat{T}_H)_2(U_H)_2- \sum_{i}N_i (A^i_H)_2^2 \right). \label{cl.pre3}
\ee
The classical prepotential which reproduces \eqref{cl.pre3} is 
\be
\mathcal{F}_{\text{classical}} = -\frac{\alpha}{4} S_H\left(\hat{T}_HU_H - \frac{1}{2}\sum_i N_i(A^i_H)^2\right). 
\ee
Similarly, the classical prepotential including generic Wilson line moduli is 
\be
\mathcal{F}_{\text{classical}} = -\frac{\alpha}{4}S_H\left(\hat{T}_HU_H - \frac{1}{2}\bar{y}^2\right) \label{cl.pre.wil}
\ee

Then, we move on to the computation of \eqref{general}. First, let us focus on the cubic terms \eqref{gauge}. Some of the cubic terms in \eqref{general} again can be absorbed by the shift $-\frac{\alpha}{4}S_H \rightarrow -\frac{\alpha}{4} + \beta U_H + \gamma \hat{T}_H + \sum_i c_iA^i_H$. A particular form of \eqref{gauge} is
\bea
(d_{\text{gauge}})_{ABC}y^Ay^By^C &=& \sum_{\bar{r}, a}d(\bar{R})\Big(\frac{(\bar{R}\cdot \bar{Q})^2}{\bar{Q}\cdot\bar{Q}}\left(\frac{2}{3}\hat{T}_H^2U_H - 4(\bar{R}\cdot\bar{y})^2\hat{T}_H\right) - \frac{1}{30}\hat{T}_H^2U_H\nonumber\\
&&+\frac{2}{3}\text{sign}(\bar{R}\cdot y_2)(\bar{R}\cdot\bar{y})^3 - \frac{1}{3}(\bar{R}\cdot\bar{y})^2U_H + \frac{1}{90}U_H^3\Big),\label{gauge.wil1}
\eea

The first line of \eqref{gauge.wil1} may be simplified further. Since \eqref{gauge.wil1} should not depend on $Q$, we choose a specific $Q$ in the $SU(16)$
\be
\bar{Q} = (1, -1, 0, \cdots, 0).
\ee
without loss of generality for the computation of the first line of \eqref{gauge.wil1}, assuming that $N_1 \geq 2$. In fact, if $N_i = 1$ for all $i$, then one cannot satisfy \eqref{req.Q}. The computation of the sum in \eqref{gauge.wil1} can be done in a similar way to the case without the Wilson line moduli. For the untwisted sector, $a=0$, the non-zero contributions come from $\bar{r} = 0$ or $\bar{r}=\text{[roots of SO(32)]}$. The sum of the first and the third term in the first line in \eqref{gauge.wil1} is 
\bea
\sum_{\bar{r}, a}d(\bar{R})\left(\frac{(\bar{R}\cdot \bar{Q})^2}{\bar{Q}\cdot\bar{Q}}\frac{2}{3}\hat{T}_H^2U_H - \frac{1}{30}\hat{T}_H^2U_H\right) &=& \frac{2}{3}(32 - 128)\hat{T}_H^2U_H - \frac{-1920}{30}\hat{T}_H^2U_H \nonumber\\
&=& 0.
\eea
On the other hand, the sum for the second term of the first line in \eqref{gauge.wil1} is 
\bea
-4\sum_{\bar{r}, a}d(\bar{R})\left(\frac{(\bar{R}\cdot \bar{Q})^2}{\bar{Q}\cdot\bar{Q}}\right)(\bar{R}\cdot\bar{y})^2\hat{T}_H &=& -4\Big\{ (8\times 2)\Big[\sum_{i=2}^kN_i(-4A^1_HA^i_H)\hat{T}_H - (N_1-2)(2A^1_H)^2\hat{T}_H\Big]\nonumber\\
&& + ((-32) \times 2 \times 2)(A^1_H)^2\hat{T}_H\Big\}\nonumber\\
&=& 0.
\eea
Hence, the sum of the first line of \eqref{gauge.wil1} in fact vanishes. Therefore, Eq.~\eqref{gauge.wil1} finally becomes
\bea
(d_{\text{gauge}})_{ABC}y^Ay^By^C &=&\sum_{\bar{r}, a}d(\bar{R})\left(\frac{2}{3}\text{sign}(\bar{R}\cdot y_2)(\bar{R}\cdot\bar{y})^3 - \frac{1}{3}(\bar{R}\cdot\bar{y})^2U_H + \frac{1}{90}U_H^3\right)\nonumber\\
&=:&\sum_{\bar{r}, a}d(\bar{R}) f(\bar{R}\cdot\bar{y}, U_H).\label{gauge.wil2}
\eea

Aiming for the comparison with \cite{Berg:2005ja}, we rewrite the sum of \eqref{gauge.wil2} in the following way, 
\bea
\sum_{\bar{r}, a=0}d(\bar{R})f(\bar{R}\cdot\bar{y}, U_H)&=&128f(0, U_H) \nonumber\\
&&+  8\sum_{1\leq i < j \leq k}N_iN_j\Big(f(A^i_H-A^j_H, U_H) + f(-A^i_H+A^j_H, U_H) \nonumber \\
&& - f(A^i_H+A^j_H, U) - f(-A^i_H-A^j_H, U_H)\Big)\nonumber \\
&& - 4\sum_{i=1}^k(N_i^2 - N_i)\left(f(2A^i_H, U_H) + f(-2A^i_H, U_H)\right)\nonumber \\
&& + 8\sum_{i=1}^k (N_i^2 - N_i)f(0, U_H)\\
&=&  4\sum_{i,j} N_iN_j\Big(f(A^i_H-A^j_H, U_H) + f(-A^i_H+A^j_H, U_H) \nonumber \\
&& - f(A^i_H+A^j_H, U_H) - f(-A^i_H-A^j_H, U_H)\Big)\nonumber \\
&& + 4\sum_{i=1}^k N_i\left(f(2A^i_H, U_H) + f(-2A^i_H, U_H)\right).\label{a0.wil}
\eea
On the other hand, the contributions from the twisted sector is 
\be
\sum_{\bar{r}, a=1}d(\bar{R})f(\bar{R}\cdot\bar{y}, U_H) = -64\sum_{i=1}^k N_i(f(A^i_H, U_H) + f(-A^i_H, U_H)), \label{a1.wil}
\ee
where we used \eqref{speciality}. Therefore, the sum of the untwisted sector \eqref{a0.wil} and the twisted sector \eqref{a1.wil} is 
\bea
\sum_{\bar{r}, a}d(\bar{R})f(\bar{R}\cdot\bar{y}, U_H) &=&  4\sum_{i,j} N_iN_j\Big(f(A^i_H-A^j_H, U_H) + f(-A^i_H+A^j_H, U_H) \nonumber \\
&& - f(A^i_H+A^j_H, U_H) - f(-A^i_H-A^j_H, U_H)\Big)\nonumber \\
&&-64\sum_{i=1}^k N_i(f(A^i_H, U_H) + f(-A^i_H, U_H)) \nonumber \\
&& + 4\sum_{i=1}^k N_i\left(f(2A^i_H, U_H) + f(-2A^i_H, U_H)\right). \label{cube.wil}
\eea

The constant term in \eqref{general} may be computed in a similar way. By summing over $a$ such that $\bar{R} \cdot \bar{y} = 0$ is satisfied for the generic value of $\bar{y}$, one arrives at
\bea
\chi &=& \frac{1}{4}\left(\frac{256}{2} + \frac{16}{2}\sum_i \frac{1}{2}(N_i^2-N_i)\times 2\right)\\
&=& 2 \sum_{i=1}^k N_i^2. \label{wil1}
\eea

The computation of the tri-logarithmic terms in \eqref{general} may be also done in a systematic way. Again we compute  first a few terms which satisfies $kl \leq 0$. For the term with $kl < 0$, there is only one term with $k=1, l=-1$ in the fundamental chamber which generates a non-zero contribution
\be
\sum_{R > 0, kl <0}d(R){\rm li}_3((R,y)) = 8{\rm Li}_3\left(e^{2\pi i(\hat{T}_H-U_H)}\right).
\ee

For the case with $kl=0$, we have two cases, (i) $k=0, ~ l > 0$ or $k>0, ~ l=0$, or (ii) $k=l=0$. In the case (i), the summation $R > 0$ involves $\bar{r}=0, \bar{r}=\text{[roots of SO(32)]}$ for $a=0$ and $-\frac{1}{2}\left(\bar{R}, \bar{R}\right) = -\frac{3}{4}$ for $a=1$. The sum is exactly the same sum of the cubic terms \eqref{cube.wil}. Hence, we have 
\bea
\sum_{\begin{subarray}{c} R > 0 \\ (k=0, l>0) \\ (k>0, l=0)\end{subarray}}d(R){\rm li}_3((R,y)) &=& \sum_{l>0} \Big\{4\sum_{i, j} N_iN_j\Big[{\rm Li}_3\left(e^{2\pi i(lU_H + A^i_H-A^j_H)}\right) + {\rm Li}_3\left(e^{2\pi i(l U_H -A^i_H+A^j_H)}\right) \nonumber \\
&&- {\rm Li}_3\left(e^{2\pi i(lU_H + A^i_H+A^j_H)}\right) - {\rm Li}_3\left(e^{2\pi i(l U_H -A^i_H-A^j_H)}\right)\Big]\nonumber \\
&&-64\sum_{i=1}^k N_i\Big[{\rm Li}_3\left(e^{2\pi i(lU_H + A^i_H)}\right) + {\rm Li}_3\left(e^{2\pi i(l U_H - A^i_H)}\right)\Big] \nonumber\\
&&+ 4\sum_{i=1}^k N_i\Big[{\rm Li}_3\left(e^{2\pi i(lU_H + 2A^i_H)}\right) + {\rm Li}_3\left(e^{2\pi i(lU_H -2A^i_H)}\right)\Big]\Big\}\nonumber\\
&&+\sum_{k>0} \Big\{4\sum_{i, j} N_iN_j\Big[{\rm Li}_3\left(e^{2\pi i(k\hat{T}_H + A^i_H-A^j_H)}\right) + {\rm Li}_3\left(e^{2\pi i(k \hat{T}_H -A^i_H+A^j_H)}\right) \nonumber \\
&&- {\rm Li}_3\left(e^{2\pi i(k\hat{T}_H + A^i_H+A^j_H)}\right) - {\rm Li}_3\left(e^{2\pi i(k\hat{T}_H -A^i_H-A^j_H)}\right)\Big]\nonumber \\
&&-64\sum_{i=1}^k N_i\Big[{\rm Li}_3\left(e^{2\pi i(k\hat{T}_H + A^i_H)}\right) + {\rm Li}_3\left(e^{2\pi i(k\hat{T}_H - A^i_H)}\right)\Big] \nonumber\\
&&+ 4\sum_{i=1}^k N_i\Big[{\rm Li}_3\left(e^{2\pi i(k\hat{T}_H + 2A^i_H)}\right) + {\rm Li}_3\left(e^{2\pi i(k\hat{T}_H -2A^i_H)}\right)\Big]\Big\}\label{wil2}
\eea

For the computation of the latter case, we need to find out the
weights which satisfy $\bar{R} > 0$ with $(\bar{R}, \bar{R}) \leq
2$. The positivity of the weights $\bar{w}$ with $(\bar{w}, \bar{w})
\leq 2$, $\bar{w} \in Spin(32)/\bbZ$ can be defined by dividing the
weights $\bar{w}$ into two sets appropriately. Since we have the
relation \eqref{speciality}, we define the positive weights as
\bea
\bar{R} &=& e_i - e_j,\;\; e_i+e_j, \qquad \text{for $a=0$},\label{positive1}\\
\bar{R} &=& 
e_m - \frac{1}{4}(e_1 + \cdots + e_{16})\qquad \text{for $a=1$}\label{positive2},
\eea
where $1\leq i < j \leq 16$ and $m=1, \cdots, 16$. However, the weights which satisfies $\bar{R}\cdot\bar{y}=0$ for generic values of $\bar{y}$ should be omitted in the sum. Therefore, we have 
\bea
\sum_{\begin{subarray}{c}R > 0 \\ k=l=0\end{subarray}}{}^{\prime}d(R){\rm li}_3((R,y)) &=& 8\sum_{1\leq i <  j \leq k}N_iN_j\Big[{\rm Li}_3\left(e^{2\pi i(A^i_H-A^j_H)}\right) - {\rm Li}_3\left(e^{2\pi i(A^i_H + A^j_H)}\right)\Big]\nonumber\\
-&4&\sum_{i=1}^k(N_i^2-N_i){\rm Li}_3\left(e^{2\pi i (2A^i_H)}\right) -64\sum_{i=1}^{k}N_i{\rm Li}_3\left(e^{2\pi i(A^i_H)}\right)\nonumber \\ 
\label{wil3}
\eea

Finally, one obtains the quantum part of the prepotential by summing up all the terms \eqref{cube.wil}, \eqref{wil1}, \eqref{wil2} and \eqref{wil3}
\bea
h(A^i_H, \hat{T}_H, U_H) &=& -i\frac{\zeta(3)\sum_i N_i^2}{4\pi ^4} -\frac{i}{2 \pi^4}{\rm Li}_3\left(e^{2\pi i(\hat{T}_H-U_H)}\right) \nonumber \\
&&- \frac{i}{16\pi^4}\Big\{8\sum_{1\leq i <  j \leq k}N_iN_j\Big[{\rm Li}_3\left(e^{2\pi i(A^i_H-A^j_H)}\right) - {\rm Li}_3\left(e^{2\pi i(A^i_H + A^j_H)}\right)\Big]\nonumber\\
&&-4\sum_{i=1}^k(N_i^2-N_i){\rm Li}_3\left(e^{2\pi i (2A^i_H)}\right) -64\sum_{i=1}^{k}N_i{\rm Li}_3\left(e^{2\pi i(A^i_H)}\right)\Big\}\nonumber\\
&&+4\sum_{i,j} N_iN_j\Big[\tilde{f}(A^i_H-A^j_H, U_H) + \tilde{f}(-A^i_H+A^j_H, U_H) \nonumber \\
&& - \tilde{f}(A^i_H+A^j_H, U_H) - \tilde{f}(-A^i_H-A^j_H, U_H)\Big]\nonumber \\
&&-64\sum_{i=1}^k N_i\Big[\tilde{f}(A^i_H, U_H) + \tilde{f}(-A^i_H, U_H)\Big] \nonumber \\
&& + 4\sum_{i=1}^k N_i\Big[\tilde{f}(2A^i_H, U_H) + \tilde{f}(-2A^i_H, U_H)\Big]\nonumber\\
&&-\frac{i}{16\pi^4}\sum_{k>0} \Big\{4\sum_{i, j} N_iN_j\Big[{\rm Li}_3\left(e^{2\pi i(k\hat{T}_H + A^i_H-A^j_H)}\right) + {\rm Li}_3\left(e^{2\pi i(k \hat{T}_H -A^i_H+A^j_H)}\right) \nonumber \\
&&- {\rm Li}_3\left(e^{2\pi i(k\hat{T}_H + A^i_H+A^j_H)}\right) - {\rm Li}_3\left(e^{2\pi i(k\hat{T}_H -A^i_H-A^j_H)}\right)\Big]\nonumber \\
&&-64\sum_{i=1}^k N_i\Big[{\rm Li}_3\left(e^{2\pi i(k\hat{T}_H + A^i_H)}\right) + {\rm Li}_3\left(e^{2\pi i(k\hat{T}_H - A^i_H)}\right)\Big] \nonumber\\
&&+ 4\sum_{i=1}^k N_i\Big[{\rm Li}_3\left(e^{2\pi i(k\hat{T}_H + 2A^i_H)}\right) + {\rm Li}_3\left(e^{2\pi i(k\hat{T}_H -2A^i_H)}\right)\Big]\Big\} + \cdots,\label{pre.wil}
\eea
where the dots in \eqref{pre.wil} stands for the contributions of the tri-logarithmic terms with $kl > 0$ and 
\be
\tilde{f}(A^i_H, U_H) := -\frac{1}{32\pi}f(A^i_H, U_H) - \frac{i}{16\pi^4}\sum_{l>0}{\rm Li}_3\left(e^{2\pi i(lU_H + A^i_H)}\right).\label{ftilde}
\ee

It is suggestive to compare \eqref{pre.wil} with the result in \cite{Berg:2005ja}. Ref.~\cite{Berg:2005ja} proposed a prepotential of a BSGP model with sixteen D9-branes and sixteen D5-branes which realize a gauge group $U(16)_9 \times U(16)_5$. Wilson lines are turned on only in a direction of the gauge group from the D9-branes and the gauge group is broken to $\Pi_i U(N_i) \times U(16)_5$ with $\sum_i N_i = 16$.  

The prepotential \eqref{pre.wil} of the heterotic string theory can be mapped to the prepotential of type I string theory by the maps \eqref{eq:I1}--\eqref{eq:I3}. For the comparison with \cite{Berg:2005ja}, we take a weak coupling limit of the string coupling $(\hat{T}_H)_2 = (\hat{S}_I^{\prime})_2 \rightarrow \infty$. 
Then, all the tri-logarithmic terms with $k > 0$ vanish in the limit and we have  
\bea
h(C^i_I, \hat{S}_I^{\prime}, U_I)|_{(\hat{S}_{I}^{\prime})_2 \rightarrow\infty} &=& -i\frac{\zeta(3)\sum_i N_i^2}{4\pi ^4}\nonumber\\
&&- \frac{i}{16\pi^4}\Big\{8\sum_{1\leq i <  j \leq k}N_iN_j\Big[{\rm Li}_3\left(e^{2\pi i(C^i_I-C^j_I)}\right) - {\rm Li}_3\left(e^{2\pi i(C^i_I + C^j_I)}\right)\Big]\nonumber\\
&&-4\sum_{i=1}^k(N_i^2-N_i){\rm Li}_3\left(e^{2\pi i (2C^i_I)}\right) -64\sum_{i=1}^{k}N_i{\rm Li}_3\left(e^{2\pi i(C^i_I)}\right) \Big\}\nonumber\\
&&+4\sum_{i, j} N_iN_j\Big[\tilde{f}(C^i_I-C^j_I, U_I) + \tilde{f}(-C^i_I+C^j_I, U_I) \nonumber \\
&& - \tilde{f}(C^i_I+C^j_I, U_I) - \tilde{f}(-C^i_I-C^j_I, U_I)\Big]\nonumber \\
&&-64\sum_{i=1}^k N_i\Big[\tilde{f}(C^i_I, U_I) + \tilde{f}(-C^i_I, U_I)\Big] \nonumber \\
&& + 4\sum_{i=1}^k N_i\Big[\tilde{f}(2C^i_I, U_I) + \tilde{f}(-2C^i_I, U_I)\Big]. \label{pre.wil1}
\eea

The prepotential \eqref{pre.wil1} is exactly the same as the proposed
prepotential in \cite{Berg:2005ja} except for the following three points. First, the second and the third line of \eqref{pre.wil1} are twice as large as the corresponding terms in \cite{Berg:2005ja}. Second, the terms 
\bea
&-&\frac{i}{32\pi^4}\Big\{8\sum_{1\leq i <  j \leq k}N_iN_j\Big[{\rm Li}_3\left(e^{2\pi i(-C^i_I+C^j_I)}\right) - {\rm Li}_3\left(e^{2\pi i(-C^i_I - C^j_I)}\right)\Big]\nonumber\\
&-&4\sum_{i=1}^k(N_i^2-N_i){\rm Li}_3\left(e^{2\pi i(-2C^i_I)}\right) -64\sum_{i=1}^{k}N_i{\rm Li}_3\left(e^{2\pi i(-C^i_I)}\right)\Big\} \label{missing}
\eea
in \cite{Berg:2005ja} are missing in \eqref{pre.wil1}. Third, the $(C^i_I)^3$ term of \eqref{gauge.wil2} does not have the sign factor,  $\text{sign}((C^i_I)_2)$, in \cite{Berg:2005ja}.\footnote{The overall constant factor of the quantum correction to the prepotential is also different
\be
-\frac{1}{8\pi}h(C^i_I, U_I) = h(C^i_I, U_I)^{\text{BHK}}.
\ee
However this is irrelevant for the physics. Indeed, playing with the ambiguity of the factor in front the axiodilaton, we can extract an overall factor in front of the full prepotential, which does not affect the low energy effective theory since it vanishes in the K\"ahler metric.}

However all the three discrepancies can be cured by taking into account the convergence of the tri-logarithmic series in \eqref{missing}. The missing terms \eqref{missing} in fact diverge in the fundamental chamber \eqref{weyl1} and \eqref{weyl2}. The tri-logarithmic series can be analytically continued outside the unit circle by the formula 
\be
{\rm Li}_3\left(e^x\right) = {\rm Li}_3\left(e^{-x}\right) + \frac{\pi^2}{3}x-\frac{i\pi}{2}x^2-\frac{1}{6}x^3. \label{analytic}
\ee
The second and the third terms in the right-hand side of \eqref{analytic} become the ambiguity in the prepotential and we can ignore them. The application of the formula \eqref{analytic} to the terms in \eqref{missing} precisely accounts for the fact that the second and the third lines of \eqref{pre.wil1} are twice as large as the corresponding terms in \cite{Berg:2005ja}. Furthermore, the term $-\frac{1}{6}x^3$ in \eqref{analytic} exactly reproduces the sign factor $\text{sign}((C^i_I)_2)$ in \eqref{ftilde}. To summarize, the prepotential \eqref{pre.wil1} precisely reproduces the prepotential in \cite{Berg:2005ja} when one takes into account the fundamental chamber \eqref{weyl1} and does the analytic continuation of the result in \cite{Berg:2005ja}.

Let us see the correspondence of the origins of the corrections on both sides. In the one-loop calculation in \cite{Berg:2005ja}, there are three types of non-zero contributions to the prepotential. First, the corrections which are proportional to $N_iN_j$ come from the one-loop diagram between the D9-branes. Second, the corrections which have a factor of $16N_i$ come from the one-loop diagram between the D5 and D9-branes. Third, the corrections which have a factor of $N_i$ and the dependence of $2C^i_I$ or $-2C^i_I$ come from the M\"obius strip diagram between the D9-branes. On the other hand, in the heterotic string theory which is dual to the BSGP model with the $SU(16)$ gauge group and sixteen half 5-branes, the first and the third types of corrections originate from the sum of the roots and the weights of the anti-symmetric representation of $SU(16)$. The second type of corrections originates from the contributions of the twisted modes at the fixed points. 

When we restrict to the case without Wilson lines, by putting $C^i_I=0$, only the second and the third types of corrections survive and sum up, as is clear from \eqref{pre.wil1}. It is easy to see that this operation gives us back the quantum corrections written in \eqref{hFunction} in the perturbative limit Im$\hat{S}^{\prime}_{I}\to\infty$ (apart from the overall factor).

\section{Duality to type IIA string compactifications}\label{DualityHet/IIA}

The $E_8 \times E_8$ heterotic string on $K3 \times T^2$ with a particular instanton embedding has a dual description of type IIA string theory on a certain Calabi--Yau threefold \cite{Kachru:1995wm, Klemm:1995tj}. The number of vector multiplets $(n_{V})$ and of hypermultiplets $(n_{H})$ arising from type IIA string theory on a Calabi--Yau threefold $X_3$ are 
\be
n_{V} = h^{1,1}(X_3), \qquad n_{H} = h^{2,1}(X_3) + 1,\label{number}
\ee
where $h^{1,1}(X_3)$ and $h^{2,1}(X_3)$ stand for the Hodge numbers of $X_3$. The gauge group of the theory is generically $U(1)^{n_{V} + 1}$ where the plus one comes from the graviphoton in a $\mathcal{N}=2$ supergravity multiplet.

Note that the plus one in the number of hypermultiplets in \eqref{number} is related to the type IIA dilaton. Hence, the vector multiplet moduli space does not receive any quantum corrections of string loops. On the other hand, the heterotic dilaton sits in a vector multiplet and the vector multiplet moduli space of the heterotic compactifications on $K3 \times T^2$  receives quantum corrections due to string loops. Therefore, the tree-level vector multiplet moduli space of the type IIA string compactifications should capture the information of string loop effects in the vector multiplet moduli space of the heterotic string compactifications. Moreover, the exact vector multiplet moduli of the type IIA string theory can be computed from the vector multiplet moduli space of type IIB string theory on the mirror Calabi--Yau threefold $\tilde{X}_3$. This is because the vector multiplet moduli space of the type IIB string compactifications on Calabi--Yau threefolds does not receive any quantum corrections, neither from $\alpha'$ nor string loops. Then, one can even study the non-perturbative effects of the vector multiplet moduli space of the heterotic compactifications from the dual type II string theory.

The vector multiplet moduli space of the type IIA string theory on $X_3$ at the large volume limit is described by the complexified K\"ahler moduli
\be\label{complexifiedJ}
B + i J = \sum_{\alpha = 1}^{h^{1,1}(X_3)}t_{\alpha} e_{\alpha},
\ee   
where $e_{\alpha}$ is a integral basis of the cohomology $H^{1,1}(X_3)$. The exact prepotential at large volume can be written as
\be
\mathcal{F}^{{\rm IIA}} = \frac{1}{6}\sum (D_{\alpha} \cdot D_{\beta} \cdot D_{\gamma})t_{\alpha}t_{\beta}t_{\gamma} + \frac{i}{(2\pi)^3}\sum_{d_1, \cdots, d_n} n_{d_1,d_2, \cdots, d_n}{\rm Li}_3(\Pi^{n}_{i=1}e^{2\pi i t_i d_i}),\label{IIA}
\ee
where $D_{\alpha}$ are the divisors associated with $e_{\alpha}$ and $n_{d_1,\cdots, d_n}$s are the rational instanton numbers. 

The first part of \eqref{IIA} is the tree-level result and the second part of \eqref{IIA} is the worldsheet instanton effects which can be computed from the mirror Calabi--Yau threefold. Apart from the non-perturbative effects, there are only cubic terms in the moduli in the prepotential. The lower order terms are just ambiguity in the prepotential and do not affect the K\"ahler metric. The higher order terms are absent by the following reason. Note that the $B$-field has a shift symmetry and this can be rephrased as the symmetry under $t_{\alpha} \rightarrow t_{\alpha}+1$. Namely, the low energy effective field theory should be invariant under the shift $t_{\alpha} \rightarrow t_{\alpha}+1$. However, if we have some terms whose orders are higher than three, then the shift symmetry generates terms whose order are higher or equal to three. Those terms alter the K\"ahler metric and indeed affect the low energy effective theory. Therefore, those terms should be absent and the perturbative prepotential contain terms whose order is up to cube. 

Let us move on to the specific examples. We considered the type \Ip string theory which is dual to
the BSGP model.
The $SO(32)$ heterotic string is dual to the $E_8 \times E_8$ heterotic string theory with the symmetric instanton embedding $(12, 12)$, and this $E_8 \times E_8$ heterotic string model has a dual type IIA model. The dual Calabi--Yau threefold $X_3$ is $W\mathbb{P}_{1,1,2,8,12}(24)$ \cite{Kachru:1995wm}. The Calabi--Yau manifold has three K\"ahler moduli which correspond to the three vector multiplet moduli $S_H, \hat{T}_H, U_H$ in heterotic compactifications. The intersection numbers of $X_3$ in a particular phase can be found in \cite{Hosono:1993qy}, for example,
\be
K(X_3) = 8J^3 - 2D^2J - 2D^2E + 8E^3,\label{int}
\ee
where $J$ is related to a divisor associated to the generating element in ${\rm Pic}(X_3)$, $D$ is related to an exceptional divisor coming from the blow-up along a singular curve, and $E$ is related to an exceptional divisor coming from the blow up at a singular point. In order to see the duality between the moduli of the type IIA compactification and the heterotic compactification, one may move to Mori's basis, which is often used in the context of mirror maps. In this case, the relation between the divisors $D_i$ associated to Mori's basis and the divisors $J, D, E$ associated to the integral basis of $H^{1,1}(X_3, \mathbb{Z})$ is \cite{Hosono:1993qy}
\be
D_1 = J + E,\quad D_2 = 2D, \quad D_3 = -D - 2E.\label{integral}
\ee
Moreover, the explicit duality maps between the K\"ahler moduli $t_i$ associated to the divisors $D_i$ and the vector multiplet moduli $S_H, T_H, U_H$ of \eqref{eq:het1}, \eqref{eq:het2}, \eqref{eq:het3} has been worked out in \cite{Klemm:1995tj} and the results are 
\bea
t_1 &=& T_H, \label{map1}\\
t_2 &=& S_H + a T_H + b U_H,\label{map2}\\
t_3 &=& U_H - T_H \label{map3}
\eea
The ambiguity in \eqref{map2} occurs since the duality map was analyzed in the weak coupling limit $(S_H)_2 \rightarrow \infty$.

With the information above, let us compute the tree-level prepotential of the type IIA compactifications on $X_3$. The classical prepotential is 
\be
\mathcal{F}^{{\rm IIA}}_{{\rm classical}} = \frac{1}{6}(J t_{J} + D t_{D} + E t_{E})^3.\label{classical}
\ee
Inserting \eqref{integral} and \eqref{map1}--\eqref{map3} into \eqref{classical}, one obtains 
\be
\mathcal{F}^{{\rm IIA}}_{{\rm classical}} = S_HT_HU_H + \frac{T_H^3}{3} + b T_H^2U_H + T_HU_H^2 + a T_HU_H^2.\label{result1}
\ee
The comparison \eqref{result1} with \eqref{HM} determines $(a, b) = (-1, 0)$. Note that the phase generating the intersection numbers \eqref{int} corresponds to the chamber $(T_H)_2 < (U_H)_2$. Furthermore, the comparison can determine the overall normalization for the prepotentials in \eqref{explicit1} or \eqref{HM}. The relevant part of the prepotential of \eqref{explicit1} is 
\be
\mathcal{F} \supset -\frac{\alpha}{4}S_H T_HU_H + \frac{2}{3\pi} T_H^3
\ee
in the region $(T_H)_2 < (U_H)_2$. Then, we choose\footnote{Here, we implicitly assume that the normalization of $S_H$ in \eqref{explicit1} and $S_H$ in \eqref{map2} is the same. This turns out to be true from the matching for the tri-logarithmic terms in the following analysis. The normalization of $S_H$ in \eqref{map2} is fixed by the relation \eqref{map2}.} the normalization $\alpha =
-\frac{8}{\pi}$, and the prepotentials on both sides are related by the overall factor
\be
\frac{\pi}{2}\mathcal{F} = \mathcal{F}^{\text{IIA}} \left( = 4\pi \mathcal{F}^{HM} \right).
\ee
In fact, the overall factor of the prepotential does not affect the low energy effective theory. From the explicit form of the K\"ahler potential \eqref{KaehlerPotPrep}, the overall factor of the prepotential becomes just the constant addition in the K\"ahler potential. Then, the constant term vanishes in the K\"ahler metric.
Note that some one-loop corrections to the prepotential of the heterotic string compactification are captured just by the tree-level computation of the type IIA string compactification. This result is indeed expected since the classical vector multiplet moduli space of the type IIA string compactification should capture string loop effects in the vector multiplet moduli space of heterotic string compactification.  

Moreover, note that the rational instanton numbers in the infinite sum of \eqref{IIA} are nothing but the genus-zero Gopakumar-Vafa invariants of the Calabi-Yau $X_3$. Indeed, $n_{d_1,d_2, \cdots, d_n}$ counts rational representatives of the class $d_\alpha e_\alpha$, which world-sheet instantons can supersymmetrically be wrapped on. This sum is supposed to reproduce the infinite series of corrections in \eqref{HM} which depend exponentially on the moduli. However, since the sum is over all possible curve classes, there will be a term corresponding to the 0-class, i.e. $d_\alpha=0$. This term will be proportional to the Euler number of the manifold, as $n_{0, \cdots, 0}$ is just enumerating points. In fact, the precise relation is \cite{RoblesLlana:2007ae}
\be
n_{0, \cdots, 0}=-\frac{\chi(X_3)}{2}\,.
\ee
The tri-logarithmic function for $d_\alpha=0$ gives rise to the Riemann zeta-function, i.e. ${\rm Li}_3(1)=\zeta(3)$. Therefore the constant term of the infinite sum in \eqref{IIA} can be written as 
\be\label{ConstantTermIIA}
\cF^{\rm IIA}_{\rm pert}=\frac{i\,\xi}{(2\pi)^3}\,,\qquad\qquad\xi\equiv-\cfrac{\chi(X_3)}{2}\,\zeta(3)\,.
\ee
Being constant, this is the only perturbative \a correction to the type IIA vector multiplet prepotential, which is compatible with the axion shift symmetry. It is the famous \a$^3$ correction, which the authors of \cite{Becker:2002nn} revisited in the $\cN=1$ context of type II orientifold compactifications. In the case at hand, i.e. $X_3=W\mathbb{P}_{1,1,2,8,12}(24)$, we have $\chi(X_3)=-480$. Dividing by $4\pi$ to get the correct normalization, we can see that \eqref{ConstantTermIIA} reproduces the constant term of \eqref{HM}.

Not only the constant term of the prepotential but also the terms of $n_{d_1,\cdots,d_n}$ with non-zero $d_{\alpha}$ should match with the tri-logarithmic terms in the prepotential of heterotic string compactifications. In other words, one may count the number of holomorphic curves by utilizing modular forms \cite{Henningson:1996jf}. Let us see this matching by comparing \eqref{IIA} in the case of $X_3=W\mathbb{P}_{1,1,2,8,12}(24)$ with \eqref{HM}. Some of the rational instanton numbers for $X_3=W\mathbb{P}_{1,1,2,8,12}(24)$ can be found in \cite{Hosono:1993qy}. The labels appearing in \cite{Hosono:1993qy} are the coefficients of the expansion in terms of $J, D, E$. Hence we denote them by $d_{J}, d_{D}, d_{E}$. By using the relation between the integral basis and the Mori basis as well as the type IIA - heterotic maps \eqref{map1}--\eqref{map3}, one can find 
\be
d_{J} = l+k,\qquad d_{D} = -k, \qquad d_{E} = l-k,
\ee
where we set $d_2 = 0$ since the duality holds only in the limit $e^{2\pi i t_2} \rightarrow 0$. Hence, we should have 
\be
-\frac{1}{2} n_{l+k, -k, l-k} = c(lk)\qquad\qquad{\rm for}\qquad kl\neq0\,, \label{duality_instanton}
\ee
where $c(h)$'s are expansion coefficients in \eqref{eisenstein}. Indeed, we can check that \eqref{duality_instanton} holds true at least for the rational instanton numbers listed in \cite{Hosono:1993qy}. This also ensures the assumption that $S_H$ in \eqref{explicit1} is the same as $S_H$ in \eqref{map2} in the previous discussion.

\bibliographystyle{JHEP}
\bibliography{refs}

\end{document}